\documentclass[a4paper]{article}

\usepackage{amsmath} 
\usepackage{amsthm} 
\usepackage{amssymb} 
\usepackage{graphicx} 
\usepackage{url}

\usepackage{multicol,lipsum}
\usepackage[latin1]{inputenc}
\usepackage{color}
\usepackage{psfrag}
\usepackage{multicol}
\usepackage[round]{natbib}
\usepackage{caption}
 
\usepackage[a4paper,total={6in,8in},bottom=1.5in,top=1in]{geometry}
\usepackage{enumitem}

\renewcommand{\cite}{\citep}

\newcommand*{\titleGM}%
{\begingroup 
\hbox{ 
\hspace*{0.05\textwidth} 
\rule{1pt}{\textheight}  
\hspace*{0.025\textwidth} 
\parbox[b]{0.80\textwidth}{ 
{\noindent\Huge\bfseries %
Degradation-Invariant 
\\}\\[0.5\baselineskip] 
{\noindent\Huge\bfseries %
Music Indexing 
\\}\\[2\baselineskip] 
{\Large \textit{ %
Onset function selection,                \\[0.25\baselineskip]
audio prints design,                     \\[0.25\baselineskip]
transformation learning,                 \\[0.25\baselineskip]
approximative hashing,                 \\[0.25\baselineskip]
search strategy.
}}
\\[6\baselineskip] 
{\large \textsc{ %
\bfseries \hspace{0.5\textwidth} Scientific Report
}} 
\\[6\baselineskip] 
{\large \textsc{ %
\bfseries R\'emi Mignot %
}} 
\\[1\baselineskip] 
{\textsc{ %
UMR 9912 STMS - IRCAM, CNRS, Sorbonne Universite, Paris, France
}}
\\[1\baselineskip] 
{\large \textsc{ %
\bfseries Geoffroy Peeters  %
}} 
\\[1\baselineskip] 
{\textsc{ %
LTCI, Telecom-Paris, Institut Polytechnique de Paris, France
}}

\vspace{0.2\textheight} 
{\noindent \today }\\[\baselineskip] 
}}
\endgroup}

\newcommand{\Rect}[2]{{R_{#2} \left( #1 \right) }}

\newcommand{\sinc}{\operatorname{sinc}}

\renewenvironment{abstract}
 {\small
  \begin{center}
  \bfseries \abstractname\vspace{-.5em}\vspace{0pt}
  \end{center}
  \list{}{%
    \setlength{\leftmargin}{15mm}
    \setlength{\rightmargin}{\leftmargin}%
  }%
  \item\relax} 
 {\endlist}

\begin{document}
\onecolumn
\titleGM
\thispagestyle{empty}
\setcounter{page}{0}

\captionsetup{width=.9\textwidth}

\begin{abstract}
For music indexing robust to sound degradations and scalable for big music catalogs, 
this scientific report presents an approach based on audio descriptors 
relevant to the music content and invariant to sound 
transformations (noise addition, distortion, lossy coding,  
pitch/time transformations, or filtering e.g.). 
To achieve this task, one of the key point of the proposed method 
is the definition of high-dimensional audio prints, 
which are intrinsically (by design) robust to some sound degradations. 
The high dimensionality of this first representation is then used 
to learn a linear projection to a sub-space significantly smaller, 
which reduces again  
the sensibility to sound degradations using a series of discriminant analyses. 
Finally, anchoring the analysis times on local maxima of a selected onset function, 
an approximative hashing is done to provide a better 
tolerance to bit corruptions, and in the same time to make easier the scaling of 
the method. 
\end{abstract}
\vspace*{1.0cm}

\section{Introduction}

\subsection{Music indexing, and sound degradations}

In audio applications, it can be useful to identify a given music 
signal, either to recognize the title of the music piece, 
or to avoid duplicates in a music catalog, for example. 
The audio indexing usually consists in representing an input sound signal 
by a sequence of audio prints\footnote{a usual name 
is \emph{audio fingerprints} because of the similarity of some methods, 
e.g. \cite{Wang2003}, with the human fingerprints recognition.}, 
then matches can be detected by comparing the codes of two signals. 
Building a database with codes of reference items, which are music pieces 
for example, it is possible to recognize a given music excerpt if 
its associated reference is included in the database. 

These audio prints are a condensed representation 
of a part of the sound, and are based on specific properties 
of the audio signals. 
However, in the case where the music excerpt to identify is 
altered by degradations, the possible modification of the audio prints 
can significantly reduce the recognition performance. 
These degradations are for example: 
noise addition, distortion or saturation, 
lossy coding/decoding (e.g. MP3, AAC, or GSM), 
pitch shifting, time stretching, equalizing, filtering, reverberation, 
dynamic range compression.
The presented work deals with the robustness of music indexing, 
that is when the signal to identify is altered by degradations, 
together with its scalability for big reference music catalogs. 

\subsection{Context of the work}

The reported work was carried out in {\bf 2015} in the context of the 
industrial project \emph{BeeMusic}. Because of possible commercial uses, 
the method has never been published yet as is. 
In consequence, this method does not take benefit of the significant advances 
of the few past years in the domain of audio processing and machine learning, 
such as the deep learning. 
It is rather based on hand-crafted descriptors, and a series of linear 
discriminant related analyses. Nevertheless, the approach is still interesting 
to present, and some points can be improved using deep learning for example. 

In order to place the following state of the art in the context of this work,
we decided to only present the known methods anterior to 2015.

\subsection{State of the art}

Most of the developed methods of the literature are splitted into two 
different steps: 
first, a signal analysis provides a sequence of audio prints which describe 
a part of the sound, ideally in a unique way. 
Second, the search step consists in finding the items of the 
reference database with the most similar audio prints. 

Among the pioneer works about audio indexing, Fraunhoffer, 
see \cite{Allamanche2001}, proposed the use of audio descriptors 
from the MPEG7 norm. 
Then, with a vector quantization (clustering), a standard classification 
approach is used to select the reference item  which is the closest 
to the given input excerpt, in the sense of a code reconstruction error. 
Each class is here defined by a reference recording. 
This method has a relative good robustness to degradations (at least 
the robustness to MP3 coding has been reported), but the method is not 
relevant for the scaling to big catalogs. 

The Philips's method of \cite{Haitsma2002} 
is based on binary codes computed with a 2D derivative filtering 
of the spectrogram. Then, after a pre-selection of candidate items, 
using a hash table, the Hamming's distance provides a 
similarity indice. 
Note that an extended version of the hash table is done, in a way that 
the search is continued until a sufficiently similar item is found. 
This principle uses a reliability indice of the bits. 
This method has a high robustness to noise addition, filtering, 
MP3 coding, and resampling, but it is reported a bad robustness to 
GSM coding and scale transformations (pitch shifting and/or 
time stretching). 

The most popular method is probably this of Shazam, 
see e.g. \cite{Wang2003}, for which the technic is based 
on audio fingerprints which represent pairs of local maxima of the 
spectrogram. 
Based on a hash table, the items of the catalogs which have the 
higher number of matching fingerprints are selected as candidates. 
Then a refinement is done by an analysis of the time-coherence 
of the matching codes (between the excerpt and the candidate items). 
This method takes advantage of the 
natural robustness of the spectral peaks to the noise addition, 
and the high number of hash codes increases the probability to find 
a sufficiently large number of unchanged codes. 
Moreover, the time-coherence makes possible to efficiently filter 
the candidate items. 
The benefit of this method is an easy scalability to big catalogs,
and an excellent robustness to the noise addition, filtering, and 
time modulation. 
But the method is not dedicated to scale transformations (pitch and time). 

To overcome to the mentioned problems of robustness, for each method, 
other works proposed extensions. 
For example, in \cite{Haitsma2003}, a transformation of the spectrum 
improved the robustness of the Philips's method to the pitch shifting. 
Moreover, in \cite{Moravec2011}, an approximative hashing has been 
introduced to this method, in order to improve the tolerance 
to bit corruptions, and so the robustness to many different sound 
degradations.
Finally let's mention \cite{Fenet2011ISMIR} 
which improves the Shazam's method for pitch shifting, 
using a Constant-Q Transform (CQT), 
and 
\cite{Sonnleitner2014quad} which uses maxima \emph{quadruplets}, 
instead of pairs, for which the temporal and frequency differences of peaks are 
coded in a relative way, in order to improve the robustness 
to both pitch and time changes.

\subsection{Categorization of the two steps}

With the studies of the previously mentioned methods, 
we remark that, in a general point of view, 
each of the two steps (1: audio prints and 2: search) 
can be splitted into two categories.  
Indeed, the audio prints describe either:  
\begin{itemize}[leftmargin=1in]
\itemsep0em 
  \item[\bf (1a)] a local information in time and frequency, or
  \item[\bf (1b)] a medium-term information, of a part of the sound. 
\end{itemize}
And the search step is based on either: 
\begin{itemize}[leftmargin=1in]
\itemsep0em 
  \item[\bf (2a)] an exact matching of the audio prints, or 
  \item[\bf (2b)] an approximative matching, e.g. computed 
               from a similarity indice. 
\end{itemize}

As said in \cite{Fenet2013PhD}, 
a good practice is to associate local audio prints with exact matching 
(1a+2a), or medium-term audio prints with approximative matching (1b+2b). 
For example the Shazam's method \cite{Wang2003}, is mainly associated to (1a+2a), 
and the Fraunhoffer's method \cite{Allamanche2001} to (1b+2b).  
Nevertheless, let's remark that the Philips's method
\cite{Haitsma2002} uses a pre-selection based on (1a+2a), 
local audio prints and hash table, and a refinement based on (1b+2b). 
Also, the refinement of the Shazam's method can be considered to be associated 
to (1b+2b), first because of the global time-coherence analysis, 
and second because it is based on the number of unchanged
hash codes, allowing few codes to be altered. 


\subsection{IrcamAudioID V1}
 
The \emph{IrcamAudioID} method, see \cite{Ramona_0} or \cite{Ramona_1}, 
is based on dedicated audio descriptors as audio prints (medium-term: 1b), 
associated to a kNN algorithm and a time-coherence analysis for the search 
step (approximative matching: 2b). 

\paragraph*{Audio print computation:}

The initial audio representation is computed by a double-nested 
\emph{Fourier Transform}, with overlapping windows of 
approximatively 2 seconds length.   
This audio descriptors were first designed in \cite{WormsMsT}, 
for audio indexing applications, and patented in \cite{RodetWormsPatent}. 
The global idea is to represent the time evolution of the signal 
characteristics, similarly to the \emph{modulation spectrum}, 
see \cite{atlas2003joint}. 

After a normalization of the signal, a standard 
\emph{Short Term Fourier Transform} (STFT) is computed
on the whole signal, and provides the representation $\left|X_{[k,\ell]}\right|$, 
with $k$ the frequency bin indice, and $\ell$ the time frame indice. 
Then, selecting a medium-term analysis window with length 2 seconds, and starting 
at $\ell=\ell_p$, a \emph{Discrete Fourier Transform} is computed over time for 
each frequency bin separately, providing the representation:
$\left|\mathcal{X}_{[k,m,p]}\right|$
with $m$ the modulation frequency indice, and $p$ the indice of the 
start time. 
Finally, the coefficients over the axes $k$ and $m$ are grouped together, 
using a weighted sum, to form a compact $(6\times6)$ representation for each 
time step $p$. 
We then obtain a sequence of $P$ vectors with size $D=36$, for all analysis times 
anchored to $\ell_p$.

\paragraph*{Search strategy and post-processing:}

The search strategy is straightforward and consists in selecting
among the code database the K-Nearest-Neighbors to the code analyzed. 
The result is a pair of matrices containing the audio item
indices and the time position in the item. 
The post-processing correlates the results of adjacent frames,
for a time-coherence analysis, 
and prunes the accidental detections of erroneous items. This procedure
discards a huge amount of elements, and when applied to sequences
of several correlated examples, only keeps the correct item indice.
This system proved to be robust to sound degradations, but 
its scaling to big catalogs is difficult because of the K-NN computation 
for each code of the excerpt. 
 
During the \emph{BeeMusic} project, 
an alternative has been proposed to make easier the scaling \cite{Regnier14_index}: 
the code sequences are converted to define words, and are used  
by the software \emph{Apache Solr}, see \cite{Solr}, in
order to index the songs. 
In return, the robustness to audio transformations 
of the whole method decreases significantly because the search step 
is then based on an exact matching (2a). 
As an application, this final system made possible the indexing 
of a catalog with more than 2 million songs (full tracks), but without 
strong degradations in accordance with the target use case.

\subsection{Overview of the method}

The present work is partly inspired from the IrcamAudioID method, 
with the aim to be both robust to sound degradations and suited to scaling. 
The following paragraphs summarize the 5 sub-steps of the method.

\paragraph*{A. Analysis times selection} ~\\
As with IrcamAudioID, see \cite{Ramona_0}, the audio prints are 
computed on an analysis window with length 2 seconds, and are associated to 
starting times $\ell_p$. 
To avoid desynchronization between the excerpt and the associated reference item, 
the analysis times are synchronized with the maxima of an onset detection function. 
Then, a \emph{maximum} filtering is processed in order to select the analysis 
times with a mean lag of $\overline{\delta_t}=0.25$ seconds.

Nevertheless, if the chosen detection function is sensitive to audio degradations, 
its maxima may be shifted in time, which would provide a new desynchronization. 
Section \ref{sec-anchoredpoints} presents some details of this initial step, 
and designing different onset functions from the literature, 
an intensive test is processed to select the best function and the best 
parameter setting, with a criterion based on the robustness. 

\paragraph*{B. High-dimensional audio print design} ~\\
Section \ref{sec-HDKey} presents the design of the descriptors used 
as high-dimensional audio prints. They are inspired by the 
\emph{modulation spectrum} of \cite{atlas2003joint}, 
but with significant differences which improve the natural robustness 
to certain sound degradations, such as the time stretching 
and the pitch shifting. 

Contrarily to the methods of Shazam and Philips, and similarly to 
the method of Fraunhoffer, these audio prints are relevant 
to the music content, by representing the evolution of the music 
characteristics over the time. 
Moreover, as mentioned above, they are robust to certain degradations 
by their design. 

\paragraph*{C. Discriminant and robust reduction} ~\\ 
The dimension of the previous representation is 1056. 
During this new step, this high-dimensional representation 
is reduced to a lower dimension of $40$ using an affine projection.  

The method takes benefit of the initial high dimension in order to 
learn the transformation with a criterion based on both the invariance
to degradation and the discrimination of different initial signals. 
Based on a data augmentation approach, a series of affine transformations 
is then learnt. 
Moreover, some transformations of the series are used to pre-condition 
the data for the following hashing. 
This step is detailed in sec.~\ref{sec-DIAPA}. 

\paragraph*{D. Hashing process}~\\ 
For the search step of the method, 
the reduced audio prints are binarized to form 40-bits integer codes, 
and are used to index a hash table, as in \cite{Wang2003,Haitsma2002}. 
This process makes possible the scalability of the method, for 
big catalogs. 

Additionally, as done in \cite{Moravec2011}, 
an approximative hashing is used 
to increase the tolerance 
to bit corruptions, and so to sound degradations. 
This step is detailed in sec.~\ref{sec-LSH}. 

\paragraph*{E. Search step and time-coherence}~\\
Based on the number of matching codes, fastly obtained through 
the hash table, a small number of items is pre-selected as 
candidates for the last sub-step. 
This final sub-step of the search, see {sec.~\ref{sec-PIE}}, 
is based on the time-coherence of the matching codes 
between the excerpt and the candidates. 
We propose here an improvement of the Shazam's method 
to deal with the time stretching transformation. 

\subsection{Outline}

First, section \ref{sec-overwiev} presents the details of all the 5 sub-steps, 
then section \ref{sec-results} presents an evaluation of the robustness, 
and finally section \ref{sec-conclusion} concludes this report. 

\newpage

\renewcommand{\thesubsection}{\thesection.\Alph{subsection}}
\section{The method}\label{sec-overwiev}

\subsection{Analysis times selection}\label{sec-anchoredpoints}
 
As with IrcamAudioID, the audio prints are computed on an analysis window 
with length 2 seconds, and are associated to starting times $\ell_p$. 
The time interval separating two adjacent analysis times has a mean of 
$\overline{\delta_t}=0.25$ seconds. 
This value has been chosen sufficiently low to obtain a fine time resolution, 
but sufficiently high to avoid a too big number of codes. 
Note that for music with a tempo of 120 BPM, a quarter of a second corresponds
to an eighth note. 

\subsubsection{Synchronization of the analysis times}

\begin{figure}[h]
  \begin{center}
    \includegraphics[width=13.0cm]{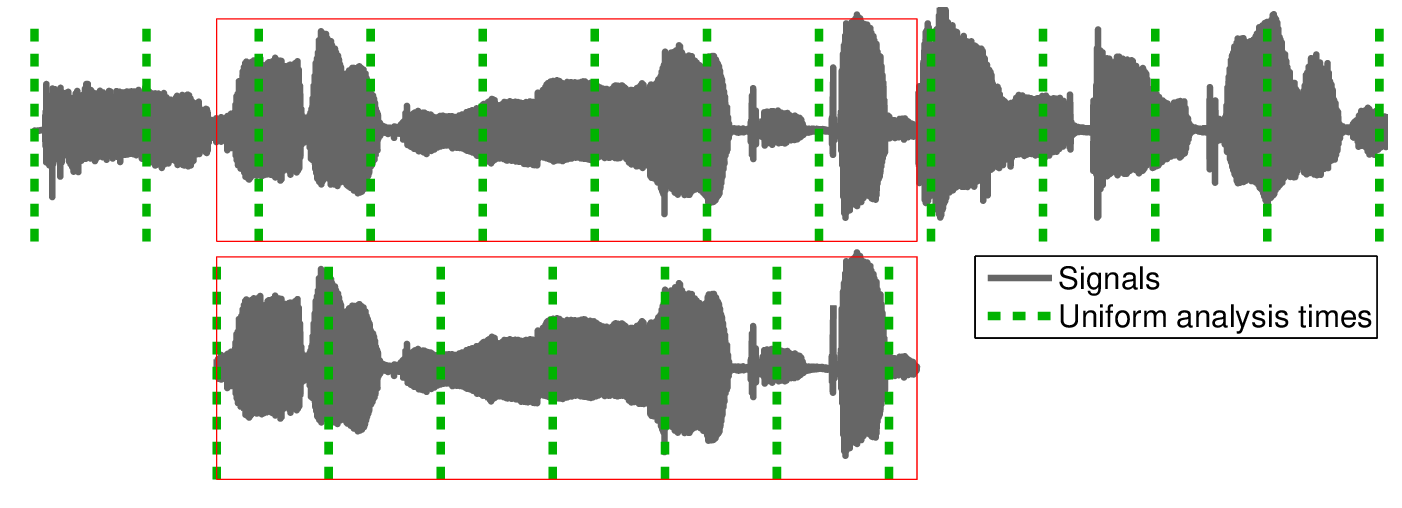}
  \end{center} 
\vspace*{-0.80cm}  
  \caption{\protect\label{fig-decalage} 
    Illustration of the desynchronization of analysis times with a uniform grid. 
    If the start point of the excerpt (below) is not synchronized on the 
    grid of the reference item (above), the analysis times 
    are misaligned (dashed vertical lines).}
\end{figure}

Nevertheless, if these times are positioned on a uniform grid, with a constant 
lag $\delta_t$, the codes could be altered in the case where the signal 
excerpt to recognize is not synchronized on the grid (see 
an illustration in Fig.~\ref{fig-decalage}).
The worst case would be with a desynchronization of $\delta_t / 2$. 
Moreover, with a time stretching transformation, 
the grids of the reference item and of the excerpt are out of sync
most of the time. 

\begin{figure}[h]
  \begin{center} 
    \includegraphics[width=13.0cm]{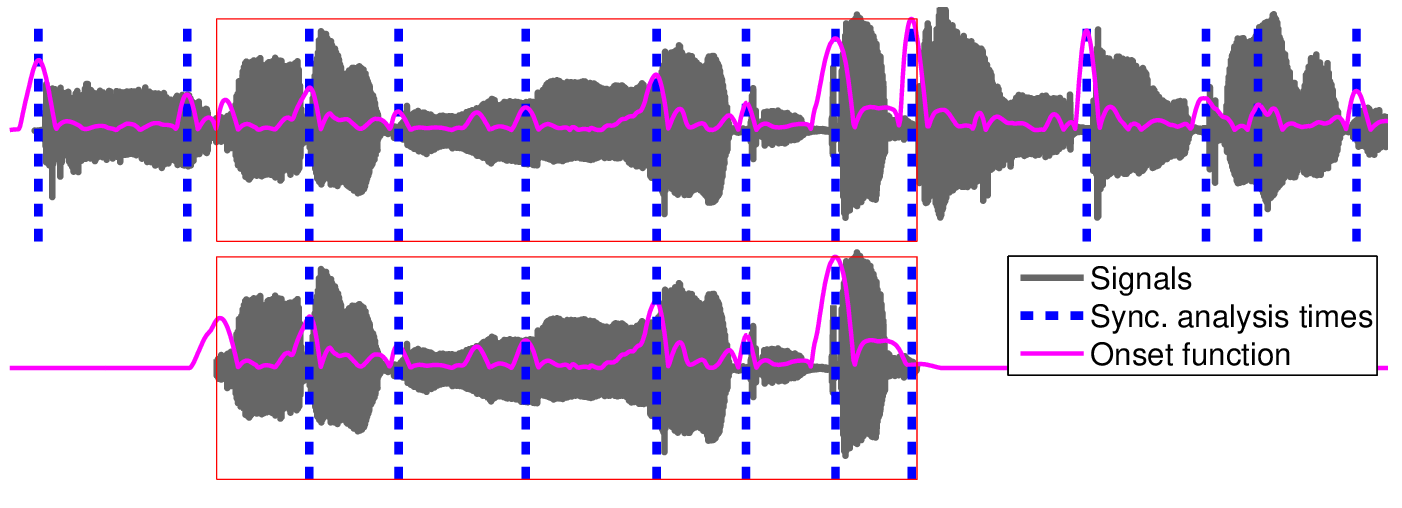} 
  \end{center}
\vspace*{-.8cm}
  \caption{\protect\label{fig-synchro} 
    Illustration of the synchronization of analysis times based 
    on an onset detection function. 
    Because the analysis times of the reference item and of the excerpt are both 
    synchronized on the onset function, the alignment is significantly 
    improved. } 
\end{figure}

To deal with this problem, in \cite{Ramona_0}, it has been proposed 
to synchronize the analysis times $\ell_p$ on the maxima of an onset detection 
function, with a mean lag $\overline{\delta_t}=0.25$ seconds. 
Onset detection functions, also known as \emph{novelty functions}, 
are functions designed with the aim to exhibit abrupt changes in the signal 
characteristics in order to detect onsets with a post-processing. 
Synchronizing the analysis times of the audio prints on the maxima 
of an onset function improves the synchronization of the codes between the 
reference item and the excerpt, whatever the starting time of the excerpt 
and the time stretching factor. 
See an illustration in Fig.~\ref{fig-synchro}.  
  
\subsubsection{Sensitivity of the onset function to degradation}

\begin{figure}[h]
  \begin{center}
    \includegraphics[width=13.0cm]{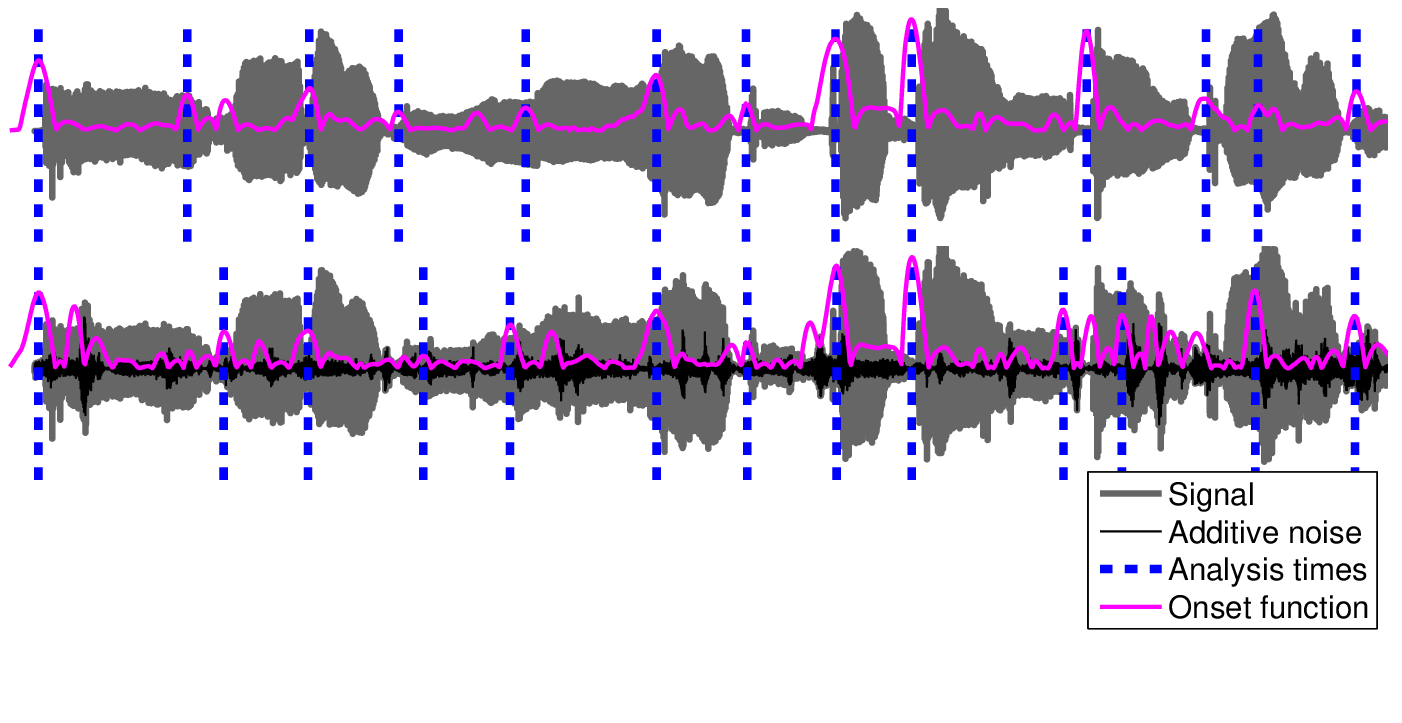}
  \end{center}
\vspace*{-1.3cm}
  \caption{\protect\label{fig-comparewithdeg}     
    Illustration of the desynchronization of analysis times with degradations.
    The results are presented for both the original signal (above)
    and its degraded version (below). With degradations, the onset function 
    is disturbed by the additive noise, and the analysis times are a shifted
    at some points.} 
\end{figure}

Unfortunately, the process may insert a new sensitivity to 
sound transformations, because the onset function may change 
with the presence of degradations of the input signal. 
See Fig.~\ref{fig-comparewithdeg} for an illustration. 
In \cite{Ramona_0} the onset function and its parameters were chosen 
without any consideration about robustness. 

\subsubsection{Towards a robust analysis time selection}\label{ssec-robustSF}
 
To improve the synchronization of the analysis with or without 
degradations, 
in the present work we have selected the onset function 
and its parameters according to criteria based on the robustness. 

For that, we have listed: 
all the parameters of the pre-processing (spectrogram computation), 
all the onset functions of the literature (and even more) 
with their parameters, 
and all the parameters of the post-filtering. 
Additionally, we tried to generalize as possible the tested onset functions, 
by inserting new parameters. 
The following paragraphs summarize the process and some parameters to set, 
and section \ref{ssec-evalSF} briefly presents the test, the chosen onset 
function and the parameter values. 
All the details of this work have been reported in \cite{RemiFluxSpectral}, 
in french. 

\paragraph*{Spectrogram} ~\\
All the tested onset functions (except one) are computed on a spectrogram (STFT) 
which is a time-frequency representation of the input sound. 
It consists in the computation of Discrete Fourier Transforms (DFT) 
on segments of the signal using a sliding and overlapping window, 
see \cite[chap. 7]{zolzerDafx2}. 
We here use a Hann window for the weighting (fixed parameter), see \cite{HarrisWindow}, 
and the sampling frequency is fixed to 11025 Hz, 
but we test some different values of the window size $w_s$, 
and the overlap step size $h_s$ in [sec] (also names \emph{hop size}). 
These parameters have an effect on the resolution of the spectrogram: 
the use of a high value of $w_s$ increases the frequency resolution 
of the spectrum but blurs the short sound events, even with a low $h_s$. 
With small values of $w_s$, and with sufficiently small $h_s$, 
the time resolution increases, but the frequency resolution decreases. 
We may think that the time resolution is better to detect onsets, 
but the true is not as straightforward because onsets can be characterized 
by smooth changes of the harmonic parts. 

Remark that with a Hann window, in most of spectral applications the following 
relation is used: $h_s \approx w_s/2$, because a higher $h_s$ implies loss of 
information, and a smaller $h_s$ implies redundancies. 
Here this relation is possibly not respected. 
Finally, converting $w_s$ in seconds to $w$ in samples, 
to use the standard FFT algorithm, each segment is zero-padded 
to get a DFT size which is the next power of 2 of $w$. 

The output spectrogram is then represented by $X_{[k,\ell]}$ 
with $k$ the frequency bin indice and $\ell$ the time frame indice.

\paragraph*{Onset functions} ~\\
For music signals, onsets are sound events in the music such as new played notes. 
They can be characterized by the attacks of the notes, or by the changes 
of the harmonic content in the case of legato playing or glissandi for examples. 
As said previously, an \emph{onset detection function} (or \emph{novelty function}) 
is a function computed on the signal (or its spectrogram) 
which exhibits changes of the signal characteristics representing 
onsets. 
In most of the onset detection methods, the result of an onset function 
is post-processed in order to pick up the time locations of the estimated 
onsets. 
In this work, the aim is not to detect onsets, by to select anchor times 
which are invariant to time shifting and sound degradations. 
Then, after the process of the onset function, a maximal filter is 
computed to extract the analysis times. 

\vspace*{0.25cm}  
To find the onset function the most robust, we tested most of 
the definitions of the literature: 
\begin{itemize}
 \item the \emph{spectral flux} (or \emph{spectral distance}), see e.g. \cite{Masri1996}, 
 \item the \emph{spectral correlation}, see e.g. \cite{Scheirer97,EssidThesis}, 
 \item the \emph{phasis deviation}, see e.g.  \cite{Bello04Phase,Duxbury2003,Dixon2006onset}, 
 \item the \emph{complex domain}, see e.g. \cite{Bello04Phase,Dixon2006onset}, 
 \item the \emph{spectral norm}, as in \cite{Ramona_1}, 
\end{itemize}

\vspace*{0.25cm}  
Also, we added new definitions of onset functions: 
\begin{itemize}
  \item the \emph{difference of spectral moments},   
  \item the \emph{difference of spectral norms}, 
  \item the \emph{difference of temporal norms}, 
  \item the \emph{Itakura-Saito divergence}, based on \cite{Itakura68}, 
  \item the \emph{Kullback-Leibler divergence}, based on \cite{KLdiv}, 
  \item the \emph{normalized Kullback-Leibler divergence}, based on \cite{Yang2011kullback}, 
  \item the \emph{generalized Kullback-Leibler divergence}, based on \cite{Yang2011kullback}, 
  \item the \emph{``Linear Prediction'' based divergence}, inspired from \cite{MakhoulLPCtuto}. 
\end{itemize}
\vspace*{0.25cm}  

Additionally, we made a generalization of the initial definitions in order 
to test more variants, and with more parameters. 
For example, with the following initial \emph{spectral flux}: 
\begin{equation}
  \left\|~ |\bf{X}_\ell| - |\bf{X}_{\ell-1}| ~\right\|_p  = 
  \begin{cases} 
    \displaystyle 
    \left( \sum_{k=1}^K  \Big| |X_{[k,\ell]}|-|X_{[k,\ell-1]}| \Big|^p \right)^{1/p} & 
      \mbox{if } p \neq +\infty,
  \vspace*{.3cm} \\ 
    \displaystyle 
    \max_{k\in[1,K]} \Big| |X_{[k,\ell]}|-|X_{[k,\ell-1]}| \Big| & 
      \mbox{if } p = +\infty.  \end{cases}
  \label{eq-normespectral}
\end{equation}
we generalized it with: 
\begin{eqnarray} 
 \phi_n^{\text{ds}}
      &=& \frac{\left\|~\Rect{|\bf{X}_\ell|-|\bf{X}_{\ell-1}|}{h}~\right\|_p}
             {\displaystyle \left(1-d\right) ~
          + ~ d \left(
                 \sqrt{\left\|\bf{X}_\ell\right\|_p   \left\|\bf{X}_{\ell-1}\right\|_p} 
  + \beta \left(~\left\|\bf{X}_\ell\right\|_p + \left\|\bf{X}_{\ell-1}\right\|_p
    \right) ~+ \epsilon ~\right) }, 
\label{eq-distancespectrale}
  \\   
  \Rect{ x }{h} &=& \big( x + h| x | \big) / ( 1+|h| )
    \label{eq-rectifh} 
\end{eqnarray}
$R_h()$ is the half-wave rectification function (with parameter $h \in [-1, 1]$), 
$\bf{X}_\ell$ is the frequency vector at the time frame $\ell$, 
$p$ is the order of the norm, 
$d \in \{0, 1\}$ is a parameter to normalize the norm, 
and $\beta \in [0, 1]$ is a mixing parameter between an algebraic or a 
geometric normalization. 
$\epsilon$ is only a small value to avoid a division by 0 with silent parts. 
See \cite{RemiFluxSpectral} for more details.

\paragraph*{Post-filtering} ~\\
To improve the robustness of the analysis time selection, 
we added a linear filtering as post-processing in order 
to smooth the time evolution.
Remark that the sampling rate of the onset functions correspond to 
the framing rate of the spectrogram which is $F_r = 1 / h_s$ Hz, 
that is the number of frames per second. 

Because the indexing is processed off-line, there is no constraint for 
real-time and for causality, then we choose a symmetrical mean-average filter. 
In consequence, the group delay is 0, and the maxima are smoothed 
and not shifted. 

We choose the window method to design the filter \cite{SASPWEB2011}: 
with $f_c$ the cutoff frequency of the low-pass filter, the time 
response $h_\ell$ of the ideal filter is 
$$ h_\ell = 2f_c  \sinc( 2f_c \ell/ F_r ), ~~ \forall \ell \in \mathbb{Z}, $$
and with $w_\ell$ a symmetrical Hamming weighting window (see \cite{HarrisWindow}), 
centered in 0 and with size $n+1$, the response of the implemented filter is
$b_\ell = h_\ell  w_\ell$, which has the support: $[-\frac{n}{2},\frac{n}{2}]$.

The tested parameters are: 
the relaxation time $t_c = 1/f_c$ in seconds, and 
the size $n$ of the filter which must be even. 
Note that a high value of $n$ may provide Gibbs effects, because of the too strong 
selectivity of the filter, and a low value may provide no smoothing. 
Finally, the input onset function $\phi_\ell$ can be modified by a non-linear 
function $\widehat{\phi_\ell} = {\phi_\ell}^r$, with $r>0$ the order. 
A value $r<1$ has the effect to compress the input function, and 
a value $r>1$ makes the maxima more predominant. 
To summarize, the post-filtering is processed using the following convolution: 
\begin{equation}
  \varphi_\ell = \Big( b ~\ast ~ {\phi}^r \Big)(\ell)
\end{equation}

\begin{figure}[h]
  \begin{center}
    \includegraphics[width=13.0cm,height=4.0cm]{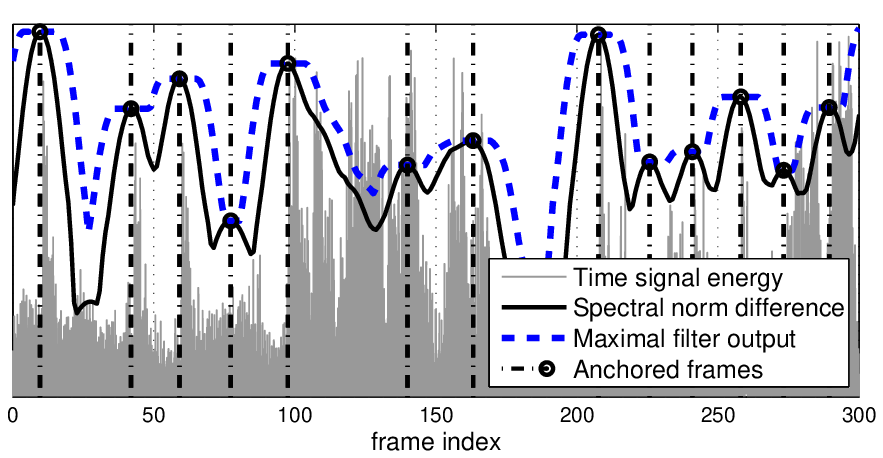}
  \vspace*{-.5cm}
  \caption{\protect\label{fig-fmax}
    Selection of the analysis times (anchored frames) 
    based on the output of the maximal filtering of the 
    smoothed onset function. In this figure, the 
    \protect\emph{spectral norm difference} is used as onset function.} 
  \end{center}
\end{figure}

\paragraph*{Maximal filtering} ~\\
For the final selection of the analysis times, a maximal filtering is used. 
This non-linear filter is processed as a median filter but extracting 
the maximal value of the signal on the sliding window rather than 
the median: 
\begin{equation}
   M_\varphi[\ell] = \max_{ m \in [-\frac{T}{2},\frac{T}{2}]} \left\{ \varphi_{\ell+m} \right\}
\end{equation}
where $T$ is the size of the window in number of frames, and it is directly linked 
to the wanted mean lag of analysis times $\overline{\delta_t} = 0.25$ 
in seconds: $ T = \overline{\delta_t} F_r$.  
As illustrated in Fig.~\ref{fig-fmax}, 
the analysis times are finally picked up by selecting the time positions 
$\ell_p$ where: $$\varphi_{\ell_p} = M_\varphi[\ell_p].$$

\subsubsection{Selection of a robust setting}\label{ssec-evalSF}

Let's remind that the goal of this preliminary work is not the detection 
of onsets, but it is the selection of analysis times robust to 
the sound degradations, and we use an onset function to achieve that. 
Then the evaluation procedure simply consisting in the comparison of the 
time positions of the selected analysis times from the original signal 
and from different degraded versions, but with the same onset function 
and associated parameters. 
As a consequence the ground truth is not ``absolute'', but it  is 
``relative'' to the parameter setting that we want to refine. 

First, the original excerpts of a music dataset have been degraded 
with a lot of different types of audio degradations: 
noise, filtering, distortion, MP3, etc. 
with different degrees of alteration, and with possible chains 
of degradations. 
This process has been reused for a data augmentation framework, 
presented in \cite{mignot2019analysis} in a music classification task. 

Then for each set of signals (original and degradations), 
the analysis times are computed for different configurations: 
onset function and parameters. 
The robustness of the configurations is then evaluated 
by comparing the analysis times obtained from the original 
signal and from the degraded signals. 
Different measures has been defined and tested, such as the ``F-measure'' 
of ``True Positives'', ``False Positives'' and ``False Negatives'', 
with a tolerance of 40ms. 

\begin{figure}[h]
  \begin{center}
    \includegraphics[width=13cm]{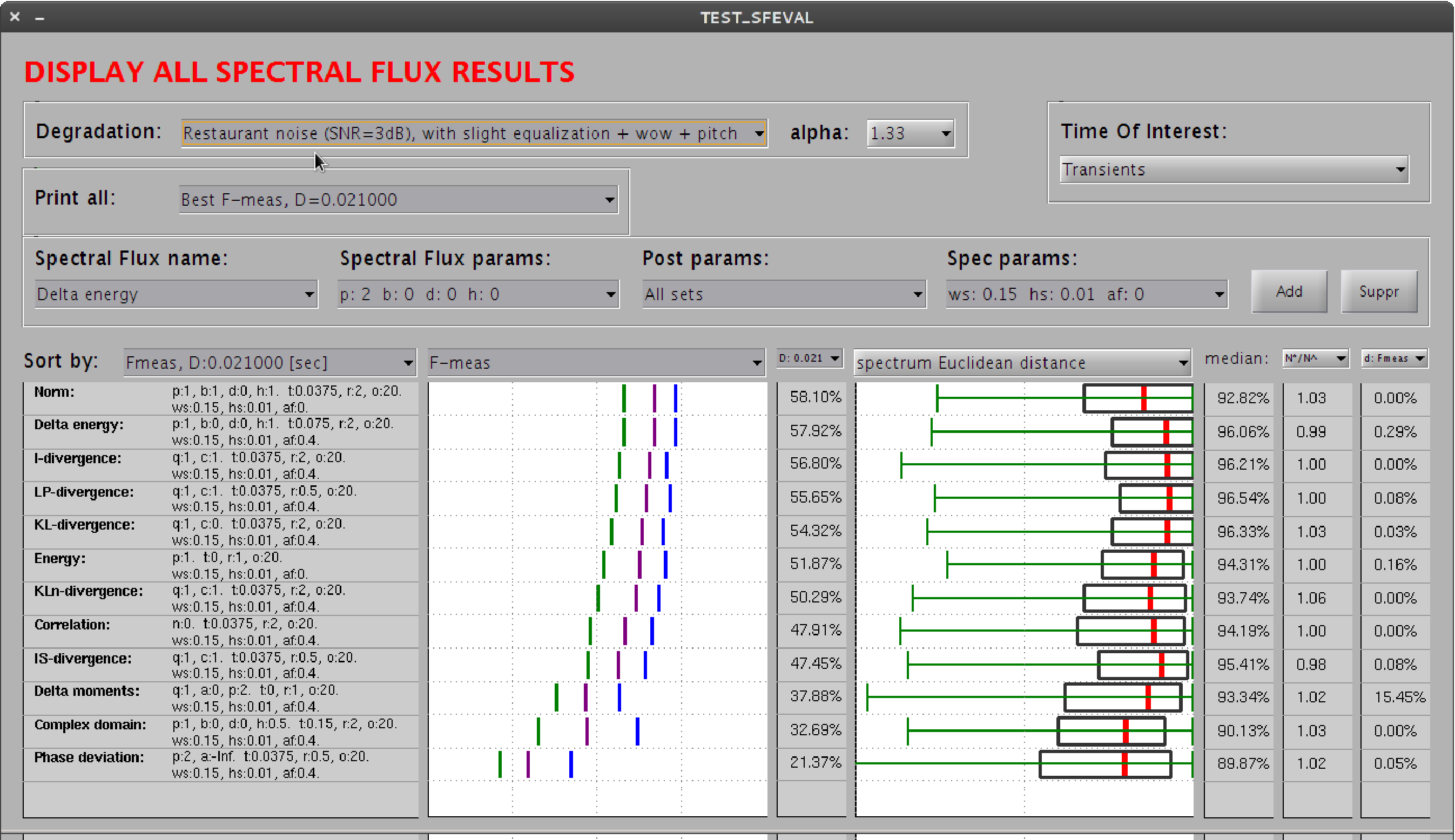}
  \end{center}
\vspace*{-.5cm}
  \caption{\protect\label{fig-GUISF}Screenshot of the graphical interface  
            which displays results.}
\end{figure}

Rather than using an automatic selection of the best configuration, 
we developed a graphical user interface to do a manual choice.
Figure \ref{fig-GUISF} presents a screenshot of this interface. 

As a result, 
two different onset functions have been ranked as the best: the first is based 
on the difference of the spectral norms, framewise, 
and the second is based on the Itakura-Saito divergence. 
For a sake of simplicity, the first one has been chosen with its parameters. 

By fixing the obtained parameters ($h=1$, $p=1$, and $d=0$), 
the obtained \emph{difference of the spectral norms} is defined by:  
\begin{equation}
  \varphi_\ell^{\text{dns}}   
    = \left|~\Rect{~\|\bf{X}_\ell\|_1 - \|\bf{X}_{\ell-1}\|_1~}{1}~\right|, 
\end{equation}
the best parameters of the spectrogram are: 
$h_s=10$ms, and $w_s = 150$ms, 
and the best parameters of the post-filtering are: 
$t_c=50$ms, $n=20$, and $r=1$.

Nevertheless, 
in the next section the computation of the high-dimensional audio prints 
is presented, and in the same way some of its parameters have been 
refined according to its robustness. 
These audio prints are also based on the computation of spectrograms, 
and the obtained parameters are slightly different. 
To use the same spectrogram computation, both for the onset function and 
for the audio prints, we decided to use the values obtained for the 
audio prints, but they do not significantly change the performances 
of the analysis times selection.
Then, the really used spectrogram parameters are: 
$h_s=20$ms, and $w_s = 150$ms, see sec.~\ref{sec-HDKey}.

\paragraph{Remark with time stretching:} ~\\ 
An important remark is about the effect of the time stretching 
on the selected analysis times. 
If the excerpt to identify is transformed by a time stretching, the time 
positions of the onsets are also stretched, and this affects the onset 
function too. 
Nevertheless, the most prominent peaks stay visible, and they are 
detected for both the excerpt and the original reference signal. 

Because the selection is set to extract an average of 4 analysis times 
per seconds, if the excerpt time is accelerated, 
some analysis times of the original signal will not be selected. 
This is not a critical problem because at least the analysis times associated 
to the more prominent peaks of the onset function are still selected. 
In the opposite case of a deceleration of the excerpt, more analysis 
times will be selected, but again it is not a critical problem. 

Let's give an example: 
a 30 seconds excerpt is extracted from a full reference recording. 
It is accelerated with a factor 1.5, and so its duration is reduced 
to 20 seconds. Whereas the analysis of the 20 seconds accelerated excerpt 
produces approximately 80 analysis times ($20\times4$), the analysis of the 
corresponding excerpt of the reference recording, 30 seconds long, produces 
120 analysis times ($30\times4$). 
Finally, among the 120 analysis times from the reference recording, 
40 do not match with the excerpt, 
but the 80 others should match because they are associated to the more prominent 
peaks of the onset function.

\subsection{High-dimensional audio print design}\label{sec-HDKey}

The developed approach permits the definition of audio prints 
which are intrinsically (by design) robust to some 
audio degradations: 
scale transformations (time and pitch), noise addition, equalizing, 
and volume modulations. 

The process of the audio prints is summarized in sec.~\ref{ssec-resumeAP}, 
and their robustness properties are given in sec.~\ref{ssec-propAP}. 
Finally, as done in sec.~\ref{sec-anchoredpoints}, 
an evaluation is briefly presented in sec.~\ref{ssec-paramAP} 
to refine the free parameters. 
More details are given in \cite{RemiHDKey}, in french.

\subsubsection{Overview of the audio prints}\label{ssec-resumeAP}

\begin{figure}[!h]
  \begin{center}
    \includegraphics
                    [width=15.0cm]{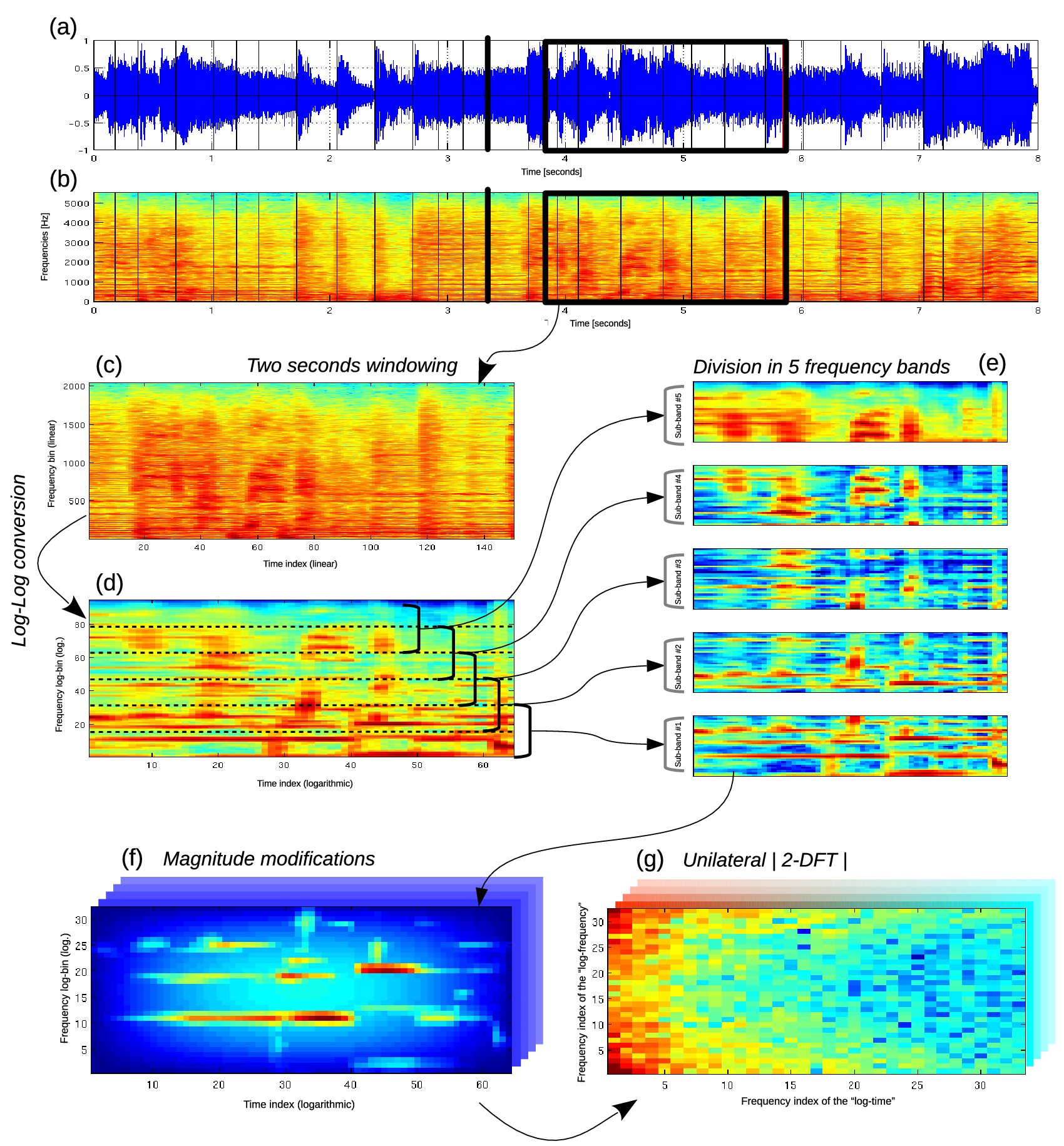}
  \end{center}
  \vspace*{-.5cm}
  \caption{\protect\label{fig-resumeHDIA}
    Illustration of the computation of the high-dimensional audio prints:
    (a) Full signal, with displayed analysis times,
    (b) Full spectrogram,
    (c) Selection of one window, synchronized on an analysis time, 
    (d) Logarithmic conversion of the frequency and the time, 
    (e) Separation into 5 frequency bands, with overlapping, 
    (f) Modification of the magnitudes: rectification, weighting,  
        normalization, log-conversion.
    (g) 2D discrete Fourier transform. } 
\end{figure}

\paragraph{1. Spectrogram} ~\\
In a first step, the whole spectrogram $\mathbb{X}_{[k,\ell]}$ 
is computed over the signal, using a \emph{Short-Time Fourier Transform}, 
see Fig.~\ref{fig-resumeHDIA}(a-b).  
The process is the same than for the analysis times selection. 
Note that with a sampling rate $S_r=11025$Hz, the frequencies are cut 
at $5500$Hz, close to the Nyquist frequency; 
the spectrogram has 2049 frequency bins.   

\paragraph{2. Medium-term analysis window} ~\\
Then a medium-term window of 3 seconds extracts a time segment of the 
whole spectrogram for the following analysis. This window is placed according 
to the analysis times $\ell_p$ selected in sec.~\ref{sec-anchoredpoints}. 
This provides the segmented spectrogram, $X_{[k,\ell]}^p = \mathbb{X}_{[k,\ell-\ell_p]}$ 
for all $\ell \in [0, L-1]$,  
see Fig.~\ref{fig-resumeHDIA}(b-c), with size $(K \times L) = (2049\times150)$: 
2049 frequency bins (unilateral spectrum), and 
150 frames. 

\paragraph{3. Log--Log conversion} ~\\
Each sub-spectrogram is a time-frequency representation 
of the signal in linear scales (seconds-hertz). 
These linear scales are then converted to logarithmic scales
both for the frequencies (log-frequencies) and the time (log-time),  
see Fig.~\ref{fig-resumeHDIA}(c-d).  
For the frequencies, it is a common process, but it is relatively 
unusual for the time. 

Note that, we need to define both a minimal frequency and a minimal time, 
because the 0 value cannot be converted to a log scale. 
The new log--log spectrogram, 
$H^p_{[ \kappa, \lambda ]}$,  
with size $(94\times64)$, represents the 
frequencies from 150 to 5000Hz, and the time from 0.5 to 2.5 seconds, 
with the associated analysis time used as reference $(t=0)$. 

This conversion is based on the Simpson's interpolation schema, 
usually used for numerical integration, and a variant has been 
developed to get good results both for over-sampling (small values of time and
frequency) and over-sampling (high values). 
$\kappa$ and $\lambda$ are then the indices of the sampled axes 
with a logarithmic profile, between the previously given limits. 

\paragraph{4. Band sub-division of the frequencies}~\\
Then, the frequency axis is splitted into 5 bands, 
see Fig.~\ref{fig-resumeHDIA}(d-e).   
These bands have an equal bandwidth in the logarithmic scale, 
and we use an overlapping factor of $1/2$. 
The obtained matrices are with size: $(32\times64)$: 
32 log-frequency bins of the band, and 64 log-time position. 
With $b \in [1,5]$ the indice of the band, the extracted matrix is: 

$$h^{p,b}_{[\kappa,\lambda]} = H^p_{[\kappa+\kappa_\text{min} ^b,\lambda ]}, 
 ~~~ \forall \kappa \in [ 0, \kappa_\text{max}^b  - \kappa_\text{min} ^b ],$$
$\kappa_\text{min} ^b$ and $\kappa_\text{max} ^b$ are the limit indices 
of the band.  
The following steps are identically processed on the 5 matrices 
of the bands.

\paragraph{5. Amplitude modification} ~\\
The amplitude values of the log--log spectrogram of each band are then 
modified. 
First, a rectification of the low values is applied: 
defining a floor value $\sigma$, adapted to $\boldsymbol{h}$, 
all magnitude below $\sigma$ are shifted to $\sigma$. 
Second, a 2D weighting is done to smoothly bring the ``pixels'' at the 
borders to 0. 
Third, a $L_\infty$ normalization is applied, 
and fourth, a quasi-logarithmic conversion of the amplitudes is done. 
The former conversion maps the range $[0,1]$ to itself, with a linear 
behavior for low values, close to 0, 
and a logarithmic behavior for values close to 1. 
See Fig.~\ref{fig-resumeHDIA}(e-f).

With $w_{[\kappa,\lambda]}$ the 2D Hamming window, $r=0.15$ and $a=10$ 
two parameters, and $\odot$ the Hadamard product, the four steps are summarized by:  
%
\begin{eqnarray}    
  &&\boldsymbol{f} = F( \boldsymbol{h} ) : \label{eq-convmag}  \\   
  && \left\{  \begin{array}{lcl} 
  \text{1) Rectification:~~~ }&
  \boldsymbol{g}  \leftarrow& \text{max} \left( \sigma , \boldsymbol{h} \right), \;\; \\
                              && \text{with} ~~ \sigma 
                                  = r ~ \text{max} \left\{\boldsymbol{h} 
                                                          \odot \boldsymbol{w} 
                                                    \right\}, \nonumber \\
  \text{2) 2D-Weighting:~~~~~ }&
  \boldsymbol{g}  \leftarrow& \boldsymbol{g} \odot \boldsymbol{w}, \\
  \text{3) Normalization:~ }&
  \boldsymbol{g}  \leftarrow& \boldsymbol{g} 
                                / \text{max} \left\{\boldsymbol{g} \right\}, \\
  \text{4) Conversion:~~~~~ }&
  \boldsymbol{f}  \leftarrow& {\log\left( 1+ a \, \boldsymbol{g}\right) } 
                                     \,/\,{\log\left(1+ a\right)}. 
\end{array}\right.
\end{eqnarray}

\paragraph{6. 2D discrete Fourier transform} ~\\
Finally, a 2D discrete Fourier transform is applied on each modified 
log--log spectrogram $\boldsymbol{h}^{p,b}$, and the modulus is extracted 
(the phasis is ignored),   
see Fig.~\ref{fig-resumeHDIA}(f-g). 
With $m$ and $n$ the indices of the frequencies of the log-frequencies
and the log-times, respectively, the final representation is: 
\begin{equation}
  \mathcal{Y}^{p,b}_{[m,n]}
    = \left| \text{2D-DFT} \left\{ \boldsymbol{f}^{p,b} \right\}_{[m,n]} \right|
\end{equation}

To reduce redundancies because of the 2D hermitian symmetry, 
we only keep one half of the obtained representation. 
Its size is then $(32\times33)$: 
32 frequency indices of the log-frequency axis, and 
33 frequency indices of the log-time axis. 
After vectorization, we obtain 1056 coefficients associated to 
one frequency band and one analysis time. 

\vspace*{0.5cm}

\paragraph{Relevancy to music content} ~\\
Contrarily to the codes of the Shazam's method, here the codes are based 
on audio prints which are relevant to the music content. 
Let's remark that, first applying a 1D-DFT over the time, instead of 
the 2D-DFT, the obtained coefficients would be similar to the modulation 
spectrum as in \cite{WormsMsT,atlas2003joint}, 
which is a Fourier representation of the evolution in time of the 
amplitude of the frequency bands. 
Second, applying a 1D-DFT over the frequency axis, 
the obtained coefficients would be roughly similar to a cepstrum or the MFCCs, 
see \cite{Davies,Rabiner}, which are related to the timbre.  
Here, the last step is a 2D-DFT (DFT both in frequency and time). 
The obtained coefficients are then meaningful for the music content, 
both for the timbre and the rhythm, they could be also used 
as descriptors for other audio applications, such as classification.  
Nevertheless, the true motivation of this process for the audio indexing 
is the quasi-invariance to both the time stretching and the pitch shifting.

\subsubsection{Invariance/robustness properties}\label{ssec-propAP}

This high-dimensional audio prints were initially designed to 
get robustness to some audio degradations. 
The next paragraphs present these properties. 

\paragraph{Quasi-invariance to scale transformations} ~\\ 
Pitch shifting and time stretching consist in the contraction or the 
expansion of the respective axis in a linear scale: frequency [Hz] 
or time [seconds]. 
In the respective logarithmic scales, such a transformation provides 
a shift of the time-frequency representation, because the $\log$ function 
transforms products to sums: $\log(at) = \log(a) + \log(t)$.
Then, a scale transform of the spectrogram $X$ in linear scales, 
see Fig.~\ref{fig-resumeHDIA}(c), becomes a translation of 
the log--log spectrogram $H$, see Fig.~\ref{fig-resumeHDIA}(d). 

Given the shift property of the Fourier transform, 
its magnitude is invariant to translation: 
$ | \text{TF}\left\{ x(t+a) \right\}| = | \text{TF}\left\{ x(t) \right\} |$, 
for all $a$ (actually the values of $a$ changes the phasis only). 
Hence, the final representation $\mathcal{Y}$, 
see Fig.~\ref{fig-resumeHDIA}(g), 
is invariant to translations of the intermediary representations, 
$\boldsymbol{H}$ and $\boldsymbol{h}$, $\boldsymbol{g}^*$ and $\boldsymbol{f}$, 
see Fig.~\ref{fig-resumeHDIA}(d,e,f), 
and by consequence to a scale transformation of $\boldsymbol{X}$, 
see Fig.~\ref{fig-resumeHDIA}(a,b,c). 

Nevertheless, in the case of discrete Fourier transforms, 
this shift property is valid for a \emph{circular} shift, 
related to the border effects, 
and this is not the case in this process. 
But, the 2D weighting of the amplitude modification reduces these border effects. 

\paragraph{Robustness to noise addition} ~\\ 
The noise is taken into at two steps in this process. 
First, considering a (quasi-)stationary noise, on an analysis window, 
and relatively localized in frequency, 
the frequency band split can isolate the effect of the noise into one or few 
bands, leaving the others unchanged. 
Second, because the noise usually has a power spectral density level 
lower than the signal of interest, the rectification of low pixel values to a 
floor $\sigma$ allows to inhibit the effect of the noise. 

Remark that a pure logarithmic behavior during the amplitude change, 
see eq. (\ref{eq-convmag}) or Fig.~\ref{fig-resumeHDIA}(e-f),
may amplify the effect of residual noises with low values, close to 0. 
The linear behavior for low values avoids possible unwanted effects.  
Remark that the limit between the linear and logarithmic behavior is defined 
by the parameter $a$, see eq. (\ref{eq-convmag}). 

\paragraph{Invariance to filtering and modulation} ~ \\
The filtering produces a product of the signal spectrum by a frequency 
function (the response of the filter) which is constant in time. 
Because of the logarithmic behavior of the amplitude change, 
see eq. (\ref{eq-convmag}) or Fig.~\ref{fig-resumeHDIA}(e-f), 
this product acts as an offset constant in time, 
see Fig.~\ref{fig-resumeHDIA}(f), and the following 2D-DFT 
isolates its effect to the first column of the final 
representation $\mathcal{Y}$. 
In consequence, all the other columns of the matrix are invariant to 
the filtering. 
In the same way, a time modulation of the amplitude 
affects the first row only. 

Then, by excluding the first row and the first column, we could 
get a representation invariant to filtering and modulations. 
Nevertheless, in this work we preserve these coefficients, 
and we leave to the dimension reduction of sec.~\ref{sec-DIAPA} 
the choice to exclude them or not if necessary. 

\paragraph{Remark on the time shift} ~\\ 
The logarithmic change of the time and frequency scales makes 
a focus on the times and frequencies close to zero (similar to an 
over-sampling). 
But a small alteration of the position of the analysis time
may produce important modifications of log--log spectrogram 
at the first indices of log-time axis. 
To reduce this unwanted effect, the minimal time indice 
of the log-time axis corresponds to 0.5 seconds after the 
position of the selected analysis time.

\subsubsection{Choice of the parameter values}\label{ssec-paramAP}

In the previous section, some parameter values are given, 
but we actually used a test to choose these values, 
as done in sec.~\ref{sec-anchoredpoints} for the analysis times. 
Even if the audio prints are designed to get a natural robustness to 
some degradations, we can expect that the robustness performance partly 
depends on the chosen values. 
These free parameters are for example the parameter of: 
the spectrogram (short-term window and hop size), 
the log--log conversion (sizes, and limits of frequencies and times), 
the number of frequency bands, 
the overlap factor of the bands, 
the shape of the 2D weighting window, 
and the parameters $a$ and $r$ of the magnitude modification, 
see eq. (\ref{eq-convmag}). 

As done in sec.~\ref{sec-anchoredpoints}, we first used a dataset
of reference excerpts which were altered with different 
audio degradations: environmental and synthetic noise addition, 
distortion, equalization, pitch shifting, wow effect (similar to a Doppler 
effect with a cyclic time-varying delay), small time shifts simulating 
small alterations of the selected analysis time positions. 
The analysis times were selected on the original excerpts, 
and reused as is on the degraded versions. 
For a sake of simplicity, we did not apply time stretching, 
but it would have been possible by an adaptation of the analysis times 
according to the stretching factor.  
Then the audio prints were computed on each excerpt (original 
and degraded versions) and for all tested configurations of parameter values. 

To evaluate the robustness performance, we defined some measures 
by a comparison of the audio prints values obtained on the original 
excerpts and the corresponding values obtained on the degraded versions, 
with the same parameter values.  
Remark that here, the ground truth depends on the tested parameter
values. 
These measures have been chosen to simulate the search process of the audio 
indexing. 

The first measure is based on the Fisher's information related to 
the separation of a \emph{positive} distribution and a \emph{negative} 
distribution. The \emph{positive} distribution is build with 
Euclidean distances between audio prints of the degraded excerpts 
with their associated original version, 
and the \emph{negative} distribution 
is build with original excerpts which do not correspond to the 
degraded version. 
Whereas the mean of the \emph{positive} distribution must be as close 
to zero as possible, the mean of the \emph{negative} distribution must 
be as high as possible. The Fisher's information gives an indice 
of separability of the two distributions. 

The second measure gives an estimate of the probability that, 
for any degraded audio print, its associated original excerpt audio print 
is among the K nearest neighbors in an Euclidean space. 
More details are given in \cite{RemiHDKey}, in french. 

\begin{figure}[h]
  \begin{center}
    \includegraphics[width=13cm]{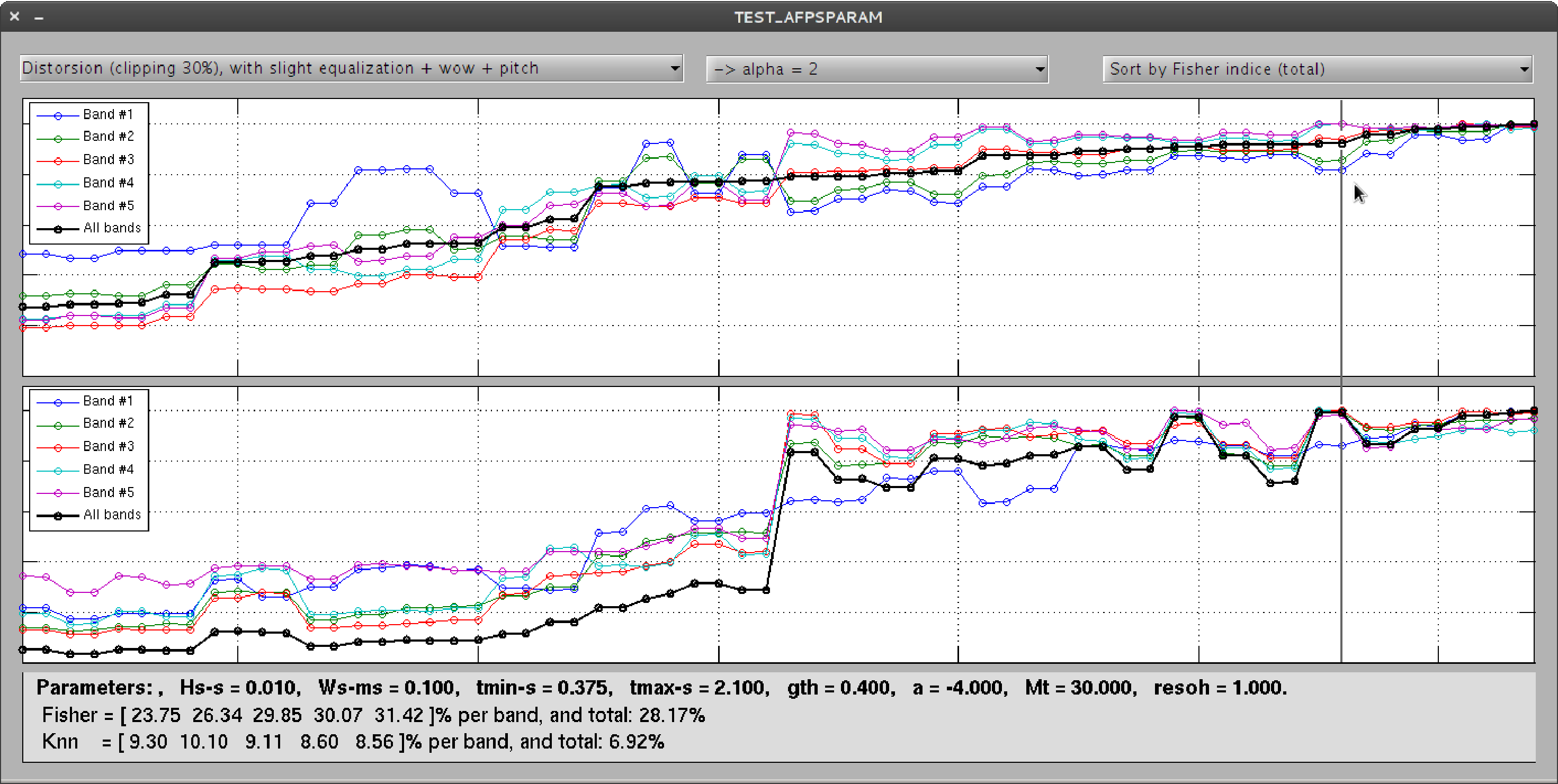}
  \end{center}
\vspace*{-.5cm}
  \caption{\protect\label{fig-GUIAP}Screenshot of the graphical interface  
            which displays results.}
\end{figure}

Instead of an automatic selection of the best value configuration, 
we developed a graphical user interface to manually 
select the configuration with a good trade-off between the two measures. 
A screenshot of the interface is given in Fig.~\ref{fig-GUIAP}. 

The final selected parameter values are the values previously given.
Note that we here obtained a spectrogram hop size of 20ms, which is not 
the best value for the analysis time selection. 
In order to compute the spectrogram only once, for both the analysis 
time selection and the audio prints, we have finally chosen 20ms. 
This does not significantly affect the robustness of the analysis time selection.

\subsection{Discriminant and robust reduction}\label{sec-DIAPA}

From the high-dimensional audio print computation, see the previous section, 
we get vectors with 1056 coefficients for all the analysis times and 
frequency bands.  
To be used with the indexing search step, this size must be reduced. 
In \cite{Ramona_0,Ramona_1}, the corresponding reduction is done by 
a weighting sum to get a new size with 36 coefficients, but without any consideration about robustness. 
 
In this work, we learn a series of affine projections with two goals: 
first the robustness, that is the reduced audio prints of degraded signal 
must be as close as possible to their corresponding original values;  
and second the discrimination, that is the reduced vectors
must be representative of the input signal. 
Indeed, the use of a constant value, independent from the input signal, 
produces of course a perfect robustness, but without any 
discrimination about the original signal. 
So, it is needed to obtain reduced audio prints which are both robust 
to audio degradations and discriminant on the original input signal. 

Moreover, some transformations of the chain pre-condition the 
variable properties to the following search step. 
The chain is built with 5 different affine or linear transformations, 
with or without reduction. Here is an overview: 
\begin{itemize}
  \item ICCR (\emph{Ill-Conditioned Component Rejection}):
    this first step guarantees that the variables used for the 
    following transformations are linearly independent. 
    This solve possible problems of matrix conditioning. 
  \item LDA (\emph{Linear Discriminant Analysis}):
    this is the first reduction with the aim to improve 
    both robustness and discrimination. 
  \item ICA (\emph{Independent Component Analysis}):
    in the search step based on hash tables, 
    to fill uniformly the buckets, one strategy is to get 
    independent variables, or at least uncorrelated. 
  \item OMPCA (\emph{Orthogonal Mahalanobis Principal Component
          Analysis}):
    the previous step provides a loss of robustness, 
    this new step has the goal to recover robustness, with a new formulation 
    closer to search process. 
  \item HT (\emph{Hadamard's Transform}):
    this last step permits to get variables with equal 
    robustness and discrimination power. 
\end{itemize}
Here is a summary of the transformation chain, with the different 
dimensions below the arrows: 
\[\begin{array}{ccccccccccc}
      \rightarrow & \text{ICCR} 
    & \rightarrow & \text{LDA} 
    & \rightarrow & \text{ICA} 
    & \rightarrow & \text{OMPCA} 
    & \rightarrow & \text{HT}  
    & \rightarrow \\ 
    1056 && 1026 && 80 && 80 && 40 && 40 
\end{array}\]

Other transformations have been tested (Data whitening, Quadratic Projections, ...), 
and also different configurations of chains. 
All the methods and some configurations have been tested using a
performance evaluation similar to these of secs.~\ref{sec-anchoredpoints} 
and \ref{sec-HDKey}. No information are given about these experiments, 
the interested reader can find details in \cite{RemiDIAPA}, in french. 
The methods and the chain which are finally used and presented in this document, 
correspond to a configuration among the best ones. 
Additionally, compared to other methods, 
this approach seems well justified. 

Note that non-linear projections have been also tested (quadratic);  
they provides interesting results, but not better. For a sake 
of simplicity, we finally selected affine or linear projections only. 

\paragraph{Separation of the bands}
An important remark is about the separate computation of the bands: 
all the following transformations are learned and applied 
on the frequency bands separately. 
The principle and the used dimensions are the same, but the learnt matrices 
depend on the band. 

\subsubsection{Ill-Conditioned Component Rejection}\label{ssec-iccr}

Even if we removed the Hermitian symmetry of the 2D-DFT, 
we know that few redundancies remain in the HD audio prints. 
The variables are then linearly dependent, and some problems may occurs 
with the matrix conditioning for the following transformations. 
We need to definitely remove these linear dependencies. 

With $J=1056$, and $X\in \mathbb{R}^J$ a random vector of HD audio prints 
computed for one time, if some of the variables of $X$ 
are linearly dependent, then a vector $g \in \mathbb{R}^J$ exists, 
such that $g^T X = \sum_j g_j X_j = 0$.  
Then the matrix $R = E( X X^T )$ similar to a covariance matrix of $X$, 
but without centering, is such that $g^T R g=0$.
In consequence, 
the eigen vectors which correspond to non-null eigen values of $R$, 
form a basis of variables which do not have any linear relation. 

Using a high number $N$ of HD audio prints, computed on different 
signals (unaltered), we build the HD audio print matrix $\bf{X}$ with 
dimensions $(J\!\times\!N)$, and with $N\gg J$. Here $N=150\,000$. 
Then the \emph{Singular Value Decomposition} (SVD) is computed: 
\begin{equation}
  {\bf X} = U S V^T,
\end{equation}
with 
$U \in \mathbb{R}^{J\!\times\!J}$ and 
$V \in \mathbb{R}^{N\!\times\!J}$ two unitary matrices, 
and 
$U \in \mathbb{R}^{J\!\times\!J}$ a diagonal matrix with singular 
values (non-negative) sorted in a decreasing order. 
Note that the columns of $V$ are mutually orthogonal. 
According to the null eigen values, we can reformulate the decomposition by: 
\begin{equation}
{\bf X} = 
\Bigg[
   ~ U_1 ~  ~ U_2 ~  
\Bigg] 
\Bigg[
\begin{array}{cc}
   S_1 &  {\bf 0}  \\  {\bf 0}  & {\bf 0} 
\end{array}
\Bigg]
\Bigg[
\begin{array}{c}
   ~~~~~~~~~~ V_1^T ~~~~~~~~~~ \\ ~~~~~~~~~~ V_2^T ~~~~~~~~~~
\end{array}
\Bigg].
\end{equation}
with $j_0$ the number of non-null singular values, 
the matrix $U_1 \in \mathbb{R}^{J\times j_0}$ provides an orthogonal 
basis of the image space of ${\bf X}$, whereas 
$V_2 \in \mathbb{R}^{N\times (J-j_0)}$ gives an orthogonal basis of the kernel 
of ${\bf X}$.
In consequence, with $P_{iccr} \triangleq U_1^T$ as projector, 
the change of variables $Z = P_{iccr} X$ removes all linear dependencies: 
\begin{equation}
  Z = P_{iccr} \, X, ~~~\text{with}~~~P_{iccr} = U_1^T.
\end{equation}

In this work, to guarantee a good conditioning of the matrices, 
we choose to define a threshold which depends on the maximal 
singular value, $\epsilon = s_1 10^{-5}$, and the selected 
vectors of $U_1$ correspond to singular values higher than $\epsilon$. 

As a result, 30 components of $X$ are linearly dependent with the other, 
which provides a reduced dimension $1026$. 
Note that some of these dependencies are explained 
by a residual symmetry between the first and last vertical columns 
of $\mathcal{Y}$, ie. at the frequencies 0 and at the Nyquist frequency 
of the log-time. 
Nevertheless we do not understand yet the other redundancies, but 
this transformation, named \emph{Ill-Conditioned Component Rejection} 
allows an automatic rejection of all linear dependencies. 

Note that, the differences of this ICCR and a PCA (\emph{Principal Component
Analysis}), are: first, a maximal number of components are kept, 
and second, the matrix $\bf{X}$ is not centered. 
Let's remark that the variables are non-negative because they 
are from the magnitude of the 2D-DFT, see sec.~\ref{sec-HDKey}.

\subsubsection{Linear Discriminant Analysis}\label{ssec-lda}

The \emph{Linear Discriminant Analysis} (LDA) is usually used 
in classification tasks. The elements of different classes are represented 
in a original vector space, and the LDA provides the optimal directions 
in order to optimally separate the associated distributions, in the sense 
of the Fisher's information, see e.g. \cite{duda2012pattern,hastie2009elements}. 

In this work, we formulated the problem of robustness for  
audio indexing in the form of a classification task. 
Each original signal is successively degraded with different audio degradations, 
and considering fixed analysis times computed on the original version, 
the HD-audio prints are computed and the ICCR is applied. 
A class distribution is then defined by all audio prints corresponding to 
the same original signal and the same analysis time, but with different  
audio degradations. 
The goal is now to estimate a reduced number of directions which 
optimally discriminate the class distributions. 

\begin{figure*}[!h]
  \begin{center}
  \includegraphics[width=10.0cm]
                    {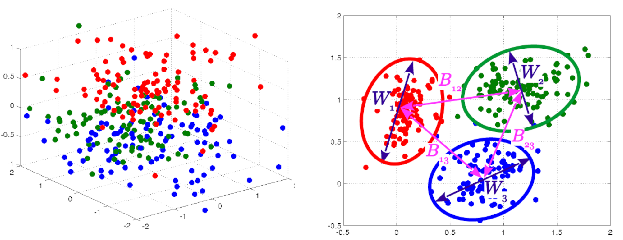}
  \end{center}
\vspace*{-.5cm}
  \caption{\protect\label{fig-LDAIllustration} 
    Illustration of an LDA. Three class distributions are represented 
    in the original 3D space on the left sub-figure, and the optimal 
    2D projection on the right. 
    Whereas the distributions look mixed in the original space, they are 
      well discriminated in the reduced space: maximizing the between-class
      covariance $B$ and minimizing the within-class covariance $W$. }
\end{figure*}

With $X$ the random variable vector of the original audio prints (with size 1026), 
and $Y$ the label of the class, the LDA roughly consists in 
maximizing the between-class covariance $ B=\text{Var}_Y[ \text{E}( X/Y) ]$
(the class centers must be kept far from each other), 
and in minimizing the within-class covariance $ W= \text{E}_Y[ \text{Var}( X/Y) ]$ 
(the class variances must be as small as possible). 
See an illustration in Fig.~\ref{fig-LDAIllustration}. 
For audio indexing, a high between-class covariance provides a discriminant 
representation of the original signals, and a small within-class covariance 
increases the robustness to audio degradations. 

For a unique variable $X_1$, the discrimination power is given by: 
$S(X_1) = \text{Var}_Y[\text{E}(X_1/Y)]/ \text{Var}(X_1)$, 
related to the Fisher's information. 
In a $J$-dimensional space, the most discriminant direction is then 
given by the vector $g\in\mathbb{R}^J$ which maximizes: 
\begin{eqnarray}
  S_g &=& \text{Var}_Y[ ~\text{E}(g^T X~/~Y)~]/ \text{Var}(g^T X), \\
      &=& ( g^T B g ) / ( g^T T g), 
\end{eqnarray}
where $T=Var[X]$ is the total covariance matrix satisfying the 
\emph{total covariance theorem}: $T = B + W$. 
Remark that this theorem implies: $0\leq S \leq 1$. 

The LDA is obtained by the diagonalization: 
$G^{-1} (T^{-1} B)G = D$, 
with $D$ a diagonal matrix of the eigen values, with size $J=1026$, 
and $G$ the invertible square matrix of the eigen vectors. 
If the values of $D$ are sorted in decreasing order, 
the LDA projector is then given by $G_{1:K}$ build with
the $K$ first eigen vectors, associated to the highest eigen values: 
\begin{equation}
  Z = P_{lda} X, ~~~\text{with} ~~~ P_{lda} = G_{1:K} ^T, 
\end{equation}
and the new components of $Z$ are ordered with decreasing power of
discrimination. 

\begin{figure*}[!h]
  \begin{center}
    \includegraphics[width=15.0cm,height=6.0cm]
                    {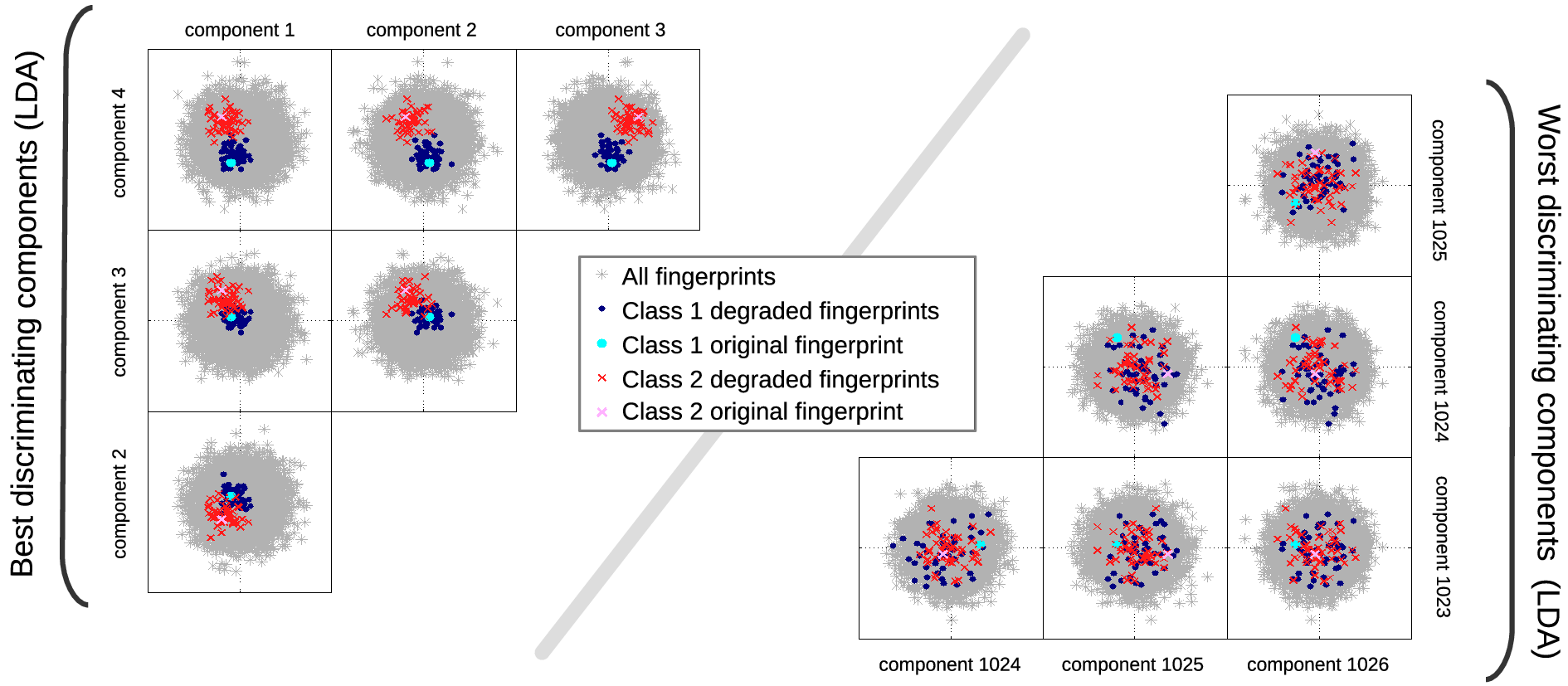}
  \end{center}
  \vspace*{-.5cm}
  \caption{\protect\label{fig-LDA-cloud} 
    Representation of the distributions of projected audio prints 
    of two classes: blue dots and red crosses. 
    The projected vectors are represented in each 
    sub-figure on 2 directions: the directions are sorted by decreasing order 
    of discrimination power, from the top-left corner (the classes are well 
    separated), to the bottom-right corner (the classes are mixed).
    The original audio prints of each class are respectively represented 
    by a cyan dot and a magenta cross. }
\end{figure*}

Remark the rank of $B$ is strictly lower than the total number classes $C$, 
which limit the choice by $K<C$. 
In this work, we used 150\,000 classes, 
built with 30\,000 musical excerpts and 5 analysis times with each. 
Each class is made of 300 degraded audio prints, and the original one. 
Since, $150\,000 \gg 1026$, we can freely choose the reduced dimension,
and we decided to select the $K=80$ most discriminating 
direction of $G$. 
Figure \ref{fig-LDA-cloud} illustrates the result. 

The use of many vectors and classes produces memory problems for the 
numerical computation of $T$ and $B$. To deal with this, the matrices 
are built in a cumulative way, by reformulating their expressions: 
\begin{eqnarray}
  T_{j,k}  &=& \frac{1}{N-1} \sum_{n=1}^N (X_j(n)-\mu_j)(X_k(n)-\mu_k), \nonumber\\
           &=& \frac{1}{N-1} \sum_{n=1}^N X_j(n) ~X_k(n)
                  - \frac{N}{N-1} \sum_{n=1}^N  \mu_j \mu_k, \\
  B_{j,k}  &=& \frac{1}{C-1} \sum_{n=1}^N (\mu_j(c)-\mu_j)(\mu_k(c)-\mu_k), \nonumber\\
           &=& \frac{1}{C-1} \sum_{n=1}^N \mu_j(c) ~ \mu_k(c)
                  - \frac{C}{C-1}  \sum_{n=1}^N  \mu_j \mu_k, 
\end{eqnarray}
with $C=150\,000$ the number of classes, 
$N=150\,000 \times 301$ the total number of vectors, 
$X_j(n)$ the $j$-th component of the $n$-th vector,  
$\mu(c)$ the mean of the class $c$, 
$\mu$ the total mean (for all the $N$ vectors), 
and then $\mu_j(c)$ and $\mu_j$ the $j$-th component of the means. 
Finally, the recursive computation is written as: 
%
 
\begin{eqnarray}
  &\begin{array}{rclcl} 
    \Gamma(n) &=& \sum_{i=1}^n X(i)         &=& \Gamma(n-1) + X(n), \\
    \Phi(n)   &=& \sum_{i=1}^n X(i) ~X(i)^T  &=& \Phi(n-1)   + X(n) ~X(n)^T, \\
    \Psi(c)   &=& \sum_{i=1}^c \mu(i) ~\mu(i)^T  &=& \Psi(c-1)   + \mu(c) ~\mu(c)^T, 
  \end{array} \nonumber\\
  &\text{with} ~~~ \mu(c) = \frac{1}{N_c} \sum_{Y(i)=c} X(i). ~~~~~~~ \nonumber \\
  &\begin{array}{rclcl} 
    \mu & = & \frac{1}{N}  \Gamma(N), \\
    T & = & \frac{1}{N-1} \Phi(N) - \frac{N}{N-1} \mu \mu^T, && ~~~~~~~~~~~~~~~~~~~~~~~\\
    B & = & \frac{1}{C-1} \Psi(C) - \frac{C}{C-1} \mu \mu^T. 
  \end{array} \nonumber\\
\end{eqnarray}

A variant has been finally tested: 
in this work, the goal is to minimize the distances between 
the degraded audio prints to their associated original versions, 
and not to the class means especially.  
In consequence, in the previous formulations, we replaced each 
class mean $\mu(c)$ by the audio print vector of the original 
associated signal. 
This provided results slightly better, but with no significant 
benefit.

\subsubsection{Independent Component Analysis}\label{ssec-ica}

During the building of the hash table, 
if some hash codes appear too many times, their associated discriminantion 
power is reduced because they carry little information.  
On the contrary, the less frequent codes are more discriminant, 
but are too rare to be really useful. 
In the previous audio indexing system, see \cite{Regnier14_index}, 
only 10\% of the hash codes were used, 
because for example many codes appeared for almost all reference songs.  

As explained in sec.~\ref{sec-LSH}, 
the components of the reduced audio prints, 
with dimension $K=40$, are binarized using their sign. 
With $z_k$ the $k$-th component of the reduced key $Z$, 
the binarization function $h$ gives the binary 
component $\gamma_k$ of the code: 
\begin{equation} \label{eq-binar}
  \gamma_k = h( z_k ) = \left\{ 
    \begin{array}{rcl}
        0, & si & z_k < 0, \\
        1, & si & z_k \geq 0.
    \end{array} \right. 
\end{equation}
The full binary hash code is then given by 
$\Gamma = ( \, \gamma_1, \, \gamma_2, \, \gamma_3, \, \dots \, \gamma_K \, )$.  
In consequence, the independence of the continuous variables $z_k$, 
or at least their decorrelation, guarantees 
the statistical independence of the binary variables $\gamma_k$.
Then all the values of $\Gamma$ between 0 and $2^{K}-1$ have the 
same probability to occur, and the buckets of the hash table (cells associated 
to hash codes) are uniformly filled. 

The goal of this step is to get new variables $Z$: 
centered around 0 ($\text{E}[Z_k]=0$ $\forall k$), 
normalized ($\text{E}[Z_k^2]=1$), 
and mutually independent, which implies the decorrelation 
$\text{E}[ZZ^T] = I$. 
Here, we use the 150\,000 audio prints computed on the original 
signals, and after the variable changes: ICCR and LDA. 
Then the Fast-ICA algorithm is used, based on the 
maximization of the neg-entropy see \cite{hyvarinen1997fastICA}. 
The method provides the following change of variables: 
\begin{equation}
  Z = P_{ica} X + t_{ica}, 
\end{equation}
with the $(J\times J)$ matrix $P_{ica}$ (where $J=80$),  
and the $(J\times 1)$ vector $t_{ICA}$ for the centering. 
As a result, we obtain a new basis of the sub-space in which the musical 
information is orthogonalized, 
and some tests proved that the hash codes occur with a more 
equal probability, as wanted. See \cite{RemiDIAPA} for more details, in french. 

Nevertheless, even if the sub-space 
is still this of the LDA, with the 80 best discriminant directions, 
we observe with an intermediary evaluation a significant loss of performance. 
It is the reason of the next step.

\subsubsection{Orthogonal Mahalanobis Principal Component Analysis}\label{ssec-ompca}

The goal of this step is the recovery of the performance, 
because of the  loss caused by the ICA. 
But in order to preserve the variable decorrelation, 
we try here to obtain an orthogonal transformation. 
Indeed, with any orthogonal matrix $P$, such that $PP^T=I$, 
and $X$ a vector of normalized, centered and mutually uncorrelated
random variables, see ICA sec.~\ref{ssec-ica}, 
the transformation $Z=PX$ preserved the mentioned properties: 
$$\text{E}[ZZ^T] = \text{E}[P X X^T P^T] 
                 = P \,\text{E}[X X^T ] \,P^T = P I P^T = P P^T = I.$$
Remark that the independence of the variables may be lost, but the more 
important property for the uniform filling of the hash table is the 
decorrelation. 
The LDA of sec.~\ref{ssec-lda} does not guarantee the orthogonality of the 
projector, we then define a new transformation 
with the same goal: improve discrimination and robustness, 
but with an orthogonal projector.
To achieve the task, first let's define two different distributions:
\begin{itemize}
 \item Each vector $\mathcal{P}_n$ of the ``\emph{Positive}'' distribution 
            is defined by 
            the difference between: $X(n)$ a degraded audio print vector, 
            and $X(c^*(n))$ the audio print vector of the corresponding 
            original signal, not altered: 
    $$ \mathcal{P}_n = X(n)-X(c^*(n)).$$
             
  \item Each vector $\mathcal{N}_n$ of the ``\emph{Negative}'' distribution 
            is defined by
            the difference between $X(n)$ a degraded audio print vector, 
            and $X(\tilde{c}(n))$ the audio print of an original signal 
            which does not correspond to $n$: 
    $$ \mathcal{N}_n = X(n)-X( \tilde{c}(n)), ~~ \text{with}~~  
            \tilde{c}(n) \neq c^*(n). $$
\end{itemize}

The goal is then to find the directions which both: 
minimize the variance of the ``\emph{positive}'' distribution in the new sub-space, 
  this informs about the error produced by degradations, 
and maximize the variance of the ``\emph{negative}'' distribution, 
  which is related to the discrimination of audio prints for different signals. 
With the covariance matrices of the two distributions,  
\newcommand{\Var}{\operatorname{Var}}
\newcommand{\E}{\operatorname{E}}
\newcommand{\pos}{\mathcal{P}}
\renewcommand{\neg}{\mathcal{N}}
\begin{eqnarray}
  \begin{array}{rcccl}
    C_\pos &\triangleq& \Var[ \pos ] &=& 
          \E\left[ ~ \left( \pos - \E(\pos) \right)  ~
                     \left( \pos - \E(\pos) \right)^T~\right], 
    \\    
    C_\neg &\triangleq& \Var[ \neg ] &=&
          \E\left[ ~ \left( \neg - \E(\neg) \right)  ~
                     \left( \neg - \E(\neg) \right)^T~\right].
    {\color{white} \Bigg(}
  \end{array}
\end{eqnarray}
the task is to maximize the variance of the \emph{negative} distribution
(similar to a PCA with $C_\neg$), 
in a vector space provided with a \emph{Mahalanobis} metric associated 
to $C_\pos$. 
We name this process: \emph{Mahalanobis Principal Component Analysis} (MPCA). 
And similarly to an LDA, this optimal directions are given by the 
diagonalization of $C_\pos^{-1} C_\neg$, given by: 
$$ G^{-1} \left( C_\pos^{-1} C_\neg \right) G = D,$$ 
with $D$ the diagonal matrix of eigen values sorted in a decreasing order, 
and $G$ an invertible matrix of the eigen vectors. 
  
Nevertheless, the orthogonality of $G$ is not guaranteed. 
A first tested solution is a Gram-Schmidt orthogonalization of 
$G$. Unfortunately we observed a new loss of performance, 
and the order of the discrimination power of the components may be lost. 

\begin{figure}[!h]
  \begin{center}
  \includegraphics[angle=90,trim=0cm 8.3cm 0cm 0cm,clip,width=14.0cm]
    {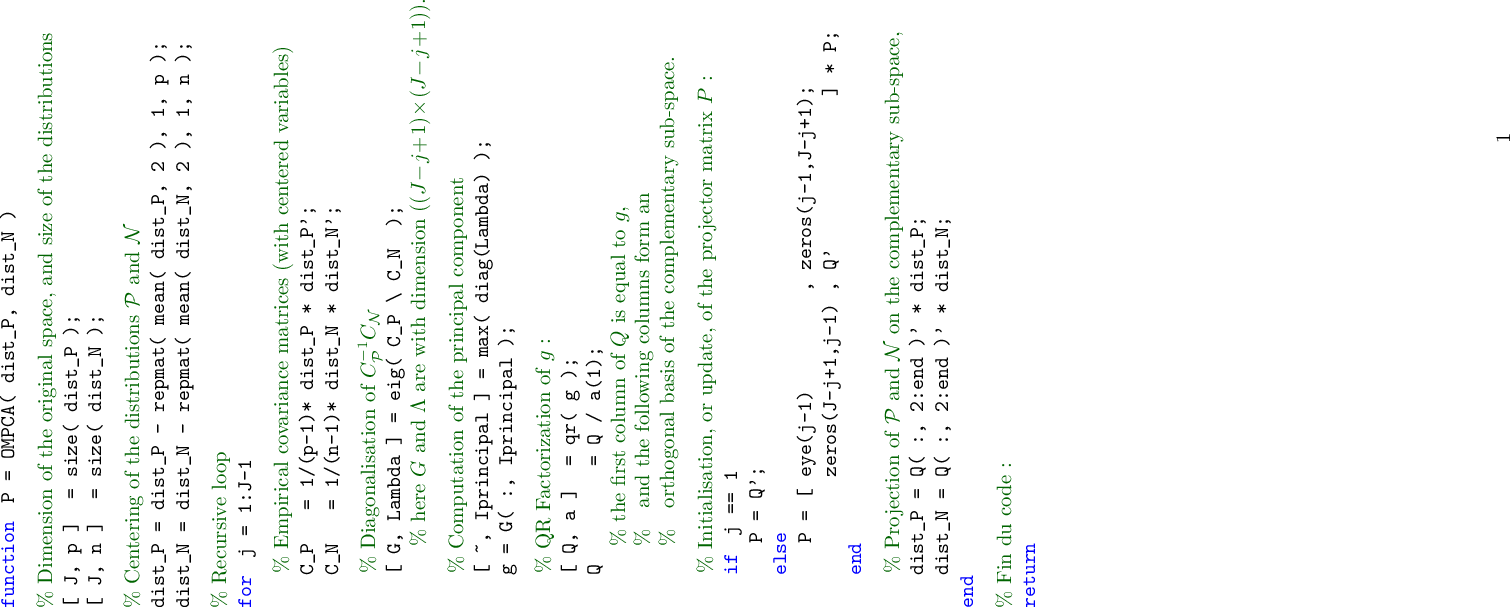}
  \end{center}  
  \caption{Matlab code of the OMPCA. 
    The two inputs are \texttt{dist\_P} and \texttt{dist\_N}
    the matrices of the two distribution: 
    $\pos$ and $\neg$. 
    The output is the orthogonal projector matrix $P$, with 
    size $(J\!\times\!J)$. 
    The selection of the $K$ first rows of $P$, provides 
    a reduction with maximal discrimination.}    \label{fig-OMPCAcode}
\end{figure}

\begin{figure}[h]
  \begin{center}
    \includegraphics[trim=2.3cm 0cm 4.5cm 0cm,clip,width=13.0cm]
                    {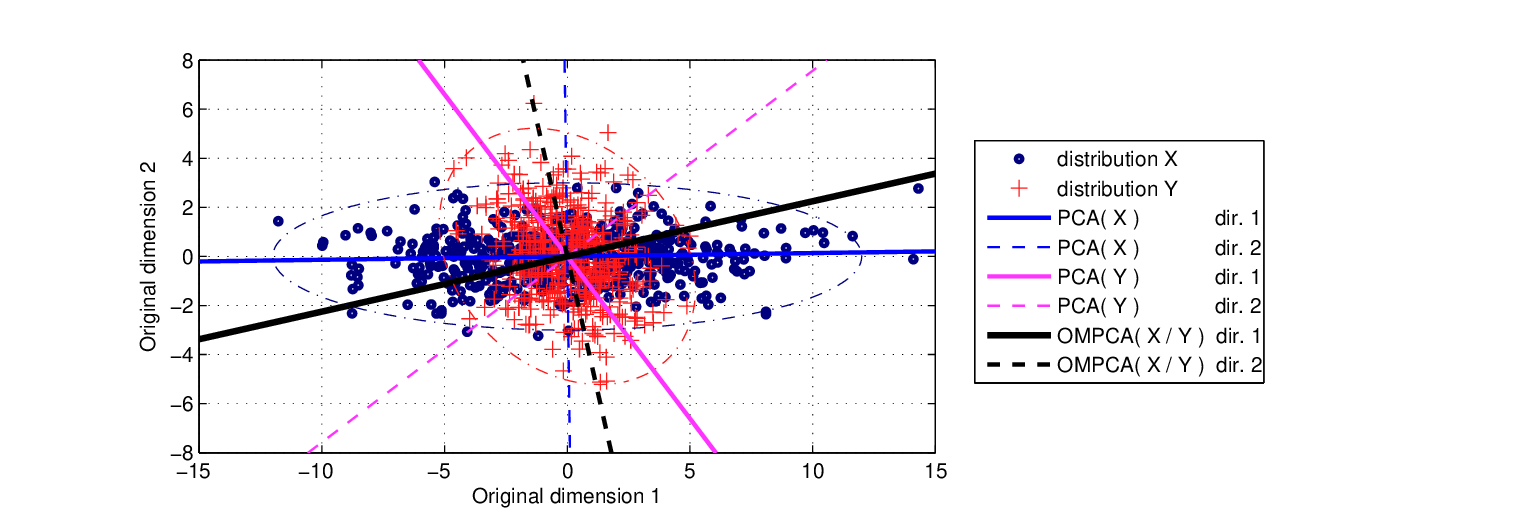}
  \end{center}
\vspace*{-.5cm}
  \caption{\protect\label{fig-OMPCA}
      PCA and OMPCA comparison.
      The PCA is applied for both $X$ and $Y$, and 
      the principal directions are displayed with plain lines 
      (maximized variance), 
      and the secondary directions with dashed lines (minimized variance). 
      For each PCA, the 2 obtained directions are orthogonal. 
      The OMPCA provides orthogonal directions, where the principal 
      direction is a trade-off between the maximization of $\Var[X]$
      and the minimization of $\Var[Y]$. }
\end{figure}

The final solution which guarantees a sorted discrimination power, 
and which provided an increased performance is 
based on an intertwined recursive computation; it can be summarized as following: 
for each step $j$, instead of projecting the $J-j$ last columns of $G$ 
on the complementary sub-space of the previous columns, 
as with the Gram-Schmidt algorithm, 
the matrices $\pos$ and $\neg$ are projected to this complementary
sub-space (with dimension $d_j=J-j$), and a new MPCA is computed. 
With $g_j$ the basis vector of the new principal component, 
a $(d_j\times d_j)$ orthogonal matrix $Q_j$ is obtained with $g_j$ as 
first column, and the global projector $P$ (with size ($J\times J$)) is updated 
by the projection of its last $d_j$ rows with $Q_j^T$. 
The process stops at the $K$-th step, with $K=40$ the dimension of 
the wanted reduced sub-space. 
For more details, Figure \ref{fig-OMPCAcode} provides a matlab 
implementation of the method;  
and Figure \ref{fig-OMPCA} illustrates the obtained directions, 
with a comparison with the PCA method, 
for a 2D problem and two distributions. 

As a result, we obtain a new linear reduction given by the following 
change of variables: 
\begin{equation}
  Z = P_{ompca} X, 
\end{equation}
with $P_{ompca}$ a ($K\times J)=(40 \times 80)$ matrix with 
mutually orthogonal rows,  
and both $X$ and $Z$ vectors of uncorrelated random variables. 
Not only this new reduction both recover the robustness and the discrimination
power, but it preserves also the decorrelation property of the 
variables: $\E[ZZ^T]=I$. 
 
Because this OMPCA step provides a discriminant reduction of dimension, 
the presence of the preliminary LDA is questionable. 
In this case, the ICA should be applied in a sub-space with dimension 
$J=1026$ and this would provides some technical problems. 
An alternative would be a so-called \emph{whitening} of the 1026 variables, 
or ZCA, which results in uncorrelated variables, as wanted, 
but different tests revealed its inefficiency to uniformly fill 
the hash table, as wanted. 
Moreover, because of the recursive process of the OMPCA, and the 
successive projection of the distribution matrices, its computation 
in a high-dimensional space (with $J=1026$) is difficult. 

In consequence, the selected sub-chain is: 
LDA $\rightarrow$ ICA $\rightarrow$ OMPCA. 
The LDA first significantly reduces the dimension from 1026 to 80, 
with an increased robustness, the ICA provides independent, and so 
uncorrelated variables, and the OMPCA recovers the robustness 
by preserving the decorrelation.

\subsubsection{Hadamard Transform}

As said, the OMPCA reduction provides variables with a 
decreasing order of robustness and discrimination. 
If this obtained reduced audio prints are binarized, as in eq. (\ref{eq-binar}), 
the robustness of the bits of the hash codes $\Gamma$ would not be equal. 
However it seems interesting to get a uniform robustness. 

In order to equalize the robustness of the audio print components 
before the binarization, we propose a last change based on the 
Hadamard transform, see e.g. \cite{kunz1979equivalence}. 
This transform is given by an orthogonal matrix where 
the modulus of its elements are equal (with real values). 
With $K$ the dimension, the modulus value is $\pm 1/\sqrt{K}$, 
and the signs are defined to get the orthogonality. 
For example, with $K = 4$:
\begin{equation}
  P^{(4)}_{hadamard} = \frac{1}{2} \left[ \begin{array}{rrrr}
      1 & 1 & 1 & 1 \\
      1 & -1 & 1 & -1 \\
      1 & 1 & -1 & -1 \\
      1 & -1 & -1 & 1
 \end{array}\right].
\end{equation}
Remark that this transform does not exist for any $K\in\mathbb{N}$. 
The first valid dimensions are: 
$$\left\{ ~1, 2, 4, 8, 12, 16, 20, 24, 32, 40, 48, 64, 80, 96,
          128, 160, 192, 256, 320, 384, 512, 640, 768, 1024 ~\right\}.$$
In this work, we use $K=40$. 
The change of variable is then given by:  
\begin{equation}
  Z = P^{(40)}_{\text{hadamard}} X. 
\end{equation}

Because each output variable $z_k$ is a linear combination of the input 
variables $x_j$, with equal coefficient magnitudes, 
the initial robustness of the $x_j$'s are equally mixed to the $z_k$'s, 
providing an equalization of the robustness. 
Moreover, because $P_{\text{hadamard}}$ is orthogonal, 
it preserves the decorrelation.

\subsubsection{Summary}

\begin{figure}[h]
  \begin{center}
    \includegraphics[width=12.0cm]{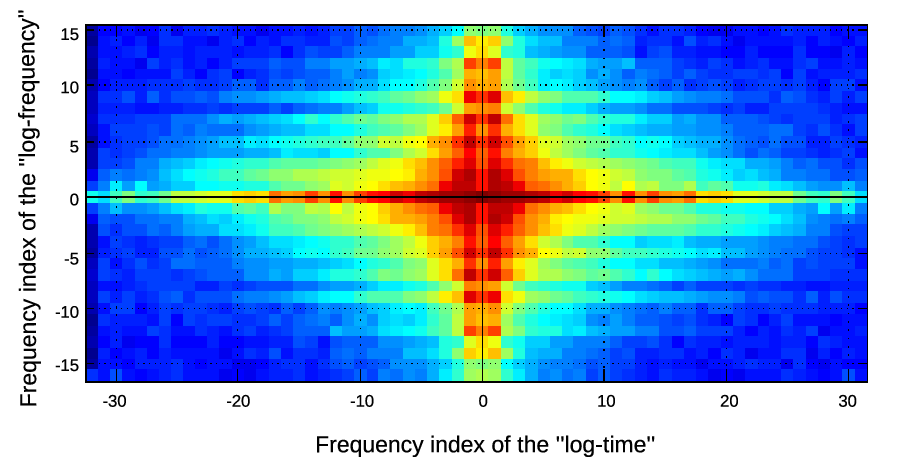}
  \end{center}
\vspace*{-.5cm} 
  \caption{\protect\label{fig-contribution}Cumulative contributions of the 
      initial components of the ($32\times33$) representation, 
      for all the $40$ reduced components (with the Hermitian symmetry).}
\end{figure}

As a result, this series of 5 affine projections return reduced audio prints 
with dimension 40, where the variables are: 
centered, normalized and mutually uncorrelated, which are important 
properties for the following hashing.  
Moreover, the most important properties are:
the learnt robustness to degradations, and their power of discrimination 
for different initial signals. 
Figure \ref{fig-contribution} presents the contributions of the 
1056 initial components. 

Remark that because all the transformation are linear or affine, 
for a faster computation, we can factorize the projection with: 
\begin{eqnarray}
  &Z = 
    {\bf P}_{\text{final}} \times X + {\bf t}_{\text{final}}, 
  &\\
  &\text{with}~~ 
  \left\{ \begin{array}{lcl}
    {\bf P}_{\text{final}} &=& 
        P_{\text{ht}} \times P_{\text{ompca}} \times 
        P_{\text{ica}} \times P_{\text{lda}} \times P_{\text{iccr}}, \\
    {\bf t}_{\text{final}} &=& P_{\text{ht}} \times P_{\text{ompca}} 
        \times  t_{\text{ica}}. 
  \end{array}\right. \nonumber
\end{eqnarray}

\paragraph*{Evaluation} ~\\
An evaluation has been done in order to test other transformations and 
other settings. As with the evaluation of the high-dimensional audio 
prints, see a summary in sec.~\ref{ssec-paramAP}, 
we used two measured which has the goal to simulate 
the next search step. 
The interested readers can find more information in \cite{RemiDIAPA}. 


\subsection{Hashing process}\label{sec-LSH}

This section summarize the hashing developed in his work. 
Not only this process allows a scaling of the method for big music catalog, 
but it also deals with bit corruptions caused by audio degradation 
using an approximative hashing. 

\subsubsection{Hash codes and hash table}

The principle of the hashing is to associate an integer to a piece of data. 
The piece of data is named the \emph{hash key}, and the obtained integer is the 
\emph{hash code} or \emph{index}. The process which computes the hash code 
from the hash key is the \emph{hash function}, 
and the \emph{hash table} acts as an \emph{inverse table} which can quickly 
provide the ids of the indexed data corresponding to a given index, 
and then we can get other information based on the obtained ids. 
To summarize the standard process of the current audio indexing, 
for each reduced audio print vector $z$ with dimension 40, 
corresponding to a song at an analysis time and a frequency band,
a hash code is computed by binarization, $\Gamma$ with 40 bits, and the 
id of the song is added to the $\Gamma$-th \emph{bucket} of the hash table. 
Let's remind that the binarization of the $k$-th component of the reduced
audio print is based on its sign, as following: 
\begin{equation} \label{eq-binar2}
  \gamma_k = h( z_k ) = \left\{ 
    \begin{array}{rcl}
        0, & si & z_k < 0, \\
        1, & si & z_k \geq 0.
    \end{array} \right. 
\end{equation}

The benefit of this approach is the speed of the answer. 
Indeed, after the computation of the HD-prints of the song, their reduction and 
their binarization, the search of matching reference songs is processed 
by extracting the id list of songs which produced the same hash codes, 
together with a histogram of the number of matching codes per reference song.  
Then, in the ideal case, the real corresponding reference may be detected based 
on the histogram. 
Note that the hash function is not bijective, that is some different 
hash keys can produce the same hash code, and so the same hash code 
can occur in different songs. Nevertheless the uniqueness of the 
identification relies on the sequence of different codes, computed at 
different analysis times and frequency bands. 
That's why a post-processing can be done in order to 
analyze the time coherence of the code occurrences between the query excerpt 
and the reference candidate songs.
\begin{figure}[h]
  \begin{center}
    \includegraphics[width=12.0cm]{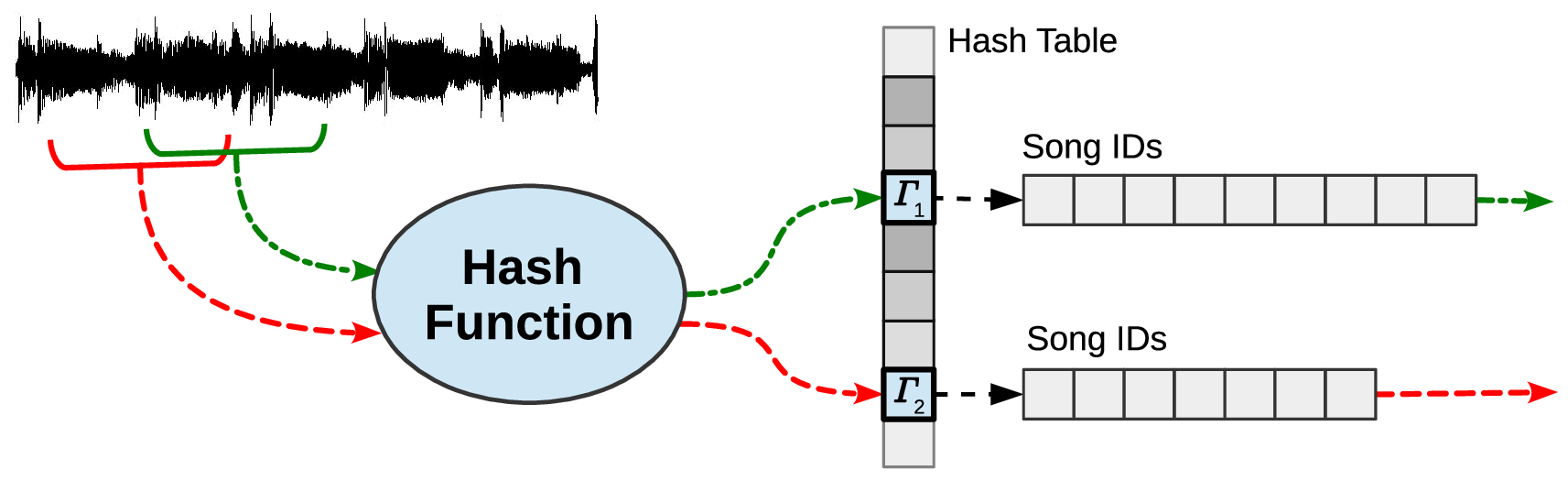}
  \end{center}
\vspace*{-.5cm}
  \caption{\protect\label{fig-hashshema}{
      Illustration of the search based on a hash function. 
      From parts of the signal (the hash keys), the hash function computes 
      a sequence of hash codes ($\Gamma_n$, for different time positions, and band). 
      Then the ids lists of the references songs are extracted from the bucket 
      indexed by $\Gamma_k$.}}
\end{figure}

\subsubsection{Approximative hashing}

Unfortunately, the process as described previously corresponds to an 
exact matching search step, and it is very sensitive to bit corruptions, 
and so to audio degradations. 
For example with audio degradations, if  at least one component $z_k$ 
changes sign (it has a high probability if the original value of $z_k$ 
is close to 0), then the value of $\gamma_k$ changes and the whole 
index $\Gamma$ is corrupted. In this case, the extracted song ids do 
not correspond to the original code. 
Even if the obtained reduced audio prints are robust to degradation and 
discriminant, this hashing process must be adapted to improve its 
tolerance to bit corruptions.  

To deal with this problem, in \cite{Moravec2011} an \emph{approximative
hashing} is proposed for audio indexing, and we developed a similar method. 
It is inspired from the \emph{Local Sensitive Hashing} (LSH) of 
\cite{Indyk1998LSH,Andoni2006LSH} which approximates the solutions of a 
problem similar to this of the \emph{K-Nearest Neighbours} algorithm, but using 
a hashing system for its speed. 
The overall idea is to derive several hash codes, using different hash functions, 
but for a same hash key. If the initial key is modified, some hash codes 
may be corrupted, but it is possible that other codes remain unchanged, 
accordingly to the strength of the modification. 

\begin{figure}[h]
  \begin{center}
    \includegraphics[width=12.0cm]{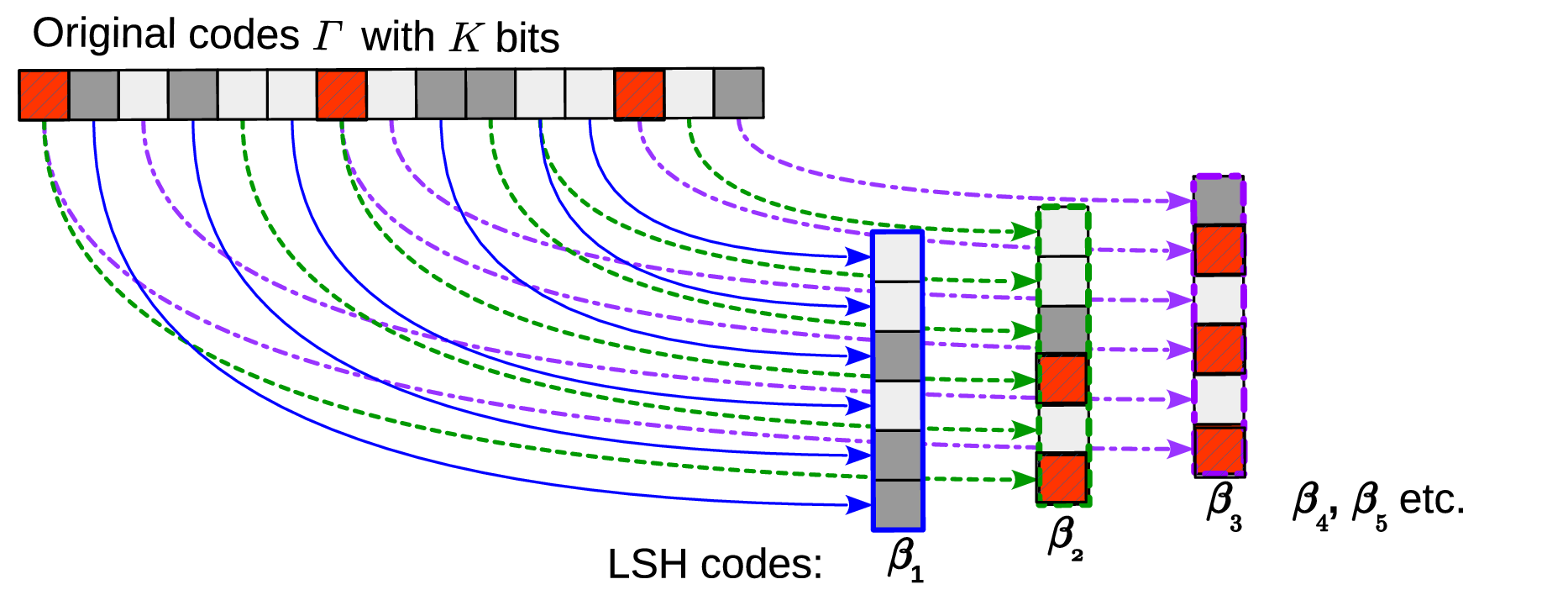}
  \end{center}
\vspace*{-.5cm}
  \caption{\protect\label{fig-lshshema}
      Illustration of the approximative hashing.
      Each lsh code $\beta_\ell$ is built by extracting 
      a selected subset of bits from $\Gamma$. 
      In this illustration, $K=15$ and $b=6$. 
      The gray cells represent bits to 1, and 
      the white cells represent bits to 0.
      The red hatched cells represent corrupted bits. 
      Whereas $\beta_2$ and $\beta_3$ are affected by the bit changes, 
      $\beta_1$ is not corrupted.}
\end{figure}

In this work, each 40 bits code $\Gamma$ is computed by binarization 
of $z$, see eq. (\ref{eq-binar2}), and $L=51$ lsh codes $\beta_\ell$ 
are computed with $b=16$ bits. 
Each smaller lsh code $\beta_\ell$ is built using 16 bits of $\Gamma$ with a
pseudo-random, and reproducible, selection. 
The principle is to use a different combination of bits from $\Gamma$ 
for each lsh codes, but the bit selections, defining the lsh functions $h_\ell$,  
remain identical for all different input $\Gamma$. 
In consequence, if some bits of $\Gamma$ are altered, 
some of the $L=51$ lsh codes are corrupted too, 
but others may be unchanged, and are still usable for the detection. 
The process is illustrated in Fig.~\ref{fig-lshshema}.

For example: with $k$ altered bits in $\Gamma$, by approximating with independent variables,
the probability that one bit is not altered becomes $p_k=(1-k/K)$, 
and the probability that a lsh code is unchanged is $P_k=p_k^b$. 
With $X$ the number of unchanged lsh codes, $0\leq X\leq L$, 
the statistical mean of unchanged lsh codes is then: 
\begin{eqnarray} \label{eq_mu}
  \mu_k = \text{E}[X ~/~ k] &=& \sum_{\ell=0}^L \ell \pi_k(\ell), \\
  \text{with}~ \pi_k(\ell) & = &  \text{Proba}[ X=\ell ~/~ k] = C_\ell^L ~ P_k^L ~ (1-P_k)^{L-\ell}. 
\end{eqnarray}
Now for unmatching audio keys, considering random drawing of $\Gamma$, 
the statistical mean of code collisions $Y$ (which does not depend on $k$, related 
to the audio degradation) is: 
\begin{equation} \label{eq_nu}
  \nu = \text{E}[Y] = L / 2^b \approx 7.782~10^{-4}. 
\end{equation}
Table \ref{tab-tableK} provides the values of $\mu_k$ and of the ratio
$\rho_k = \mu_k / \nu$ 
for $0\leq k \leq 20$. Note that because $K=40$, with $k=20$, that is half of 
bits are corrupted, the ratio is 1: as many good matches as there 
are code collisions with unmatching audio keys. 
\begin{table}[h]
\centering
  \begin{tabular}{ | r || c|c|c|c|c|c|c|c|c|c|c|c|c|c|c|c|c|c|c|c|c | }
  \hline 
  $k$ & 
  0     & 1  & 2  & 3  & 4    & 5    & 6    & 7  & 8   & 9       \\ 
  \hline 
  $\mu$ & 51 &  34.0  & 22.4 &  14.6   & 9.45 &    6.02 &  3.78  & 2.34 & 1.43 &    0.86 \\ 
  \hline 
  $\rho$ &65536 & 43.7\,e3 & 28.8\,e3 & 18.8\,e3 & 12.1\,e3 & 7.7\,e3 & 4.8\,e3 & 3.0\,e3 
          & 1.8\,e3 & 1.1\,e3 \\ 
  \hline  
  \end{tabular}
  \\ ~ \\~ \\ 
  \begin{tabular}{ | r || c|c|c|c|c|c|c|c|c|c|c|c|c|c|c|c|c|c|c|c|c | }
  \hline 
  $k$ & 10 & 11 &  12  &   13 & 14 & 15 & 16 & 17 & 18 & 19 & 20 \\ 
  \hline 
  $\mu$ & 0.51 & 0.29  & 0.16 & 0.095 & 0.052 & 0.028 & 0.014 & 0.007 & 0.004 & 0.002 & 0.001 \\ 
  \hline 
  $\rho$ & 656.8 & 381.8 &217.8 & 121.7 & 66.5 & 35.5 & 18.5 & 9.35 & 4.59 & 2.18 & 1 \\
  \hline 
  \end{tabular}
  \caption{\protect\label{tab-tableK} $\mu_k$: statistical mean of unchanged lsh codes 
      with $k$ altered bits (in $\Gamma$), 
      and $\rho_k$: ratio $\mu_k/\nu$, with $\nu$ the statistical mean 
      of collisions (for unmatching audio keys).
      The values of $\mu$ and $\rho$ are rounded to the most significant digits.}
\end{table}


\subsubsection{For a unique hash table}

For each analysis time, $L$ lsh codes (with $b=16$ bits)
are obtained for all the $n_b=5$ frequency bands. 
The $L \times n_b$ obtained codes cannot be mixed to index the same hash table 
with size $2^b$. 
Either we built $L \times n_b$ different hash tables, 
or equivalently we can extend the hash table with $S = L \times n_b \times  2^b$ 
entries, or buckets. 
With $n_b=5$, and $L=51$, the number of lsh codes for each analysis times 
is then $n_b \times L = 255$ which can be coded with 8 bits.   
Then, to simplify the implementation, we build an extended hash table, 
which is indexed with extended codes on $24$ bits.

\subsubsection{Lightening of the table using the most reliable codes}

To make lighter hash tables, one idea is to limit the number of used 
lsh codes for each analysis time and each band. 
That is, instead of using all the $L$ obtained lsh codes, only $L'<L$ are used. 
Among the $L$ codes, the idea is to choose these ones which are the less 
sensible to degradations. Based on a reliability indice as in \cite{Haitsma2002}, 
the used lsh codes are the ones which have the smallest probability 
to be altered.  

To explain the principle, let's take the example of an audio print component 
$z_k=0$. With eq.~(\ref{eq-binar2}), its binarization produces $\gamma_k=1$. 
But adding a ``centered'' random value to $z_k$, simulating the degradation effect, 
the probability that $\gamma_k$ becomes 0 is $0.5$. 
In the opposite case, the further $z_k$ is from 0, 
the more the probability of change of $\gamma_k$ decreases.
In consequence, for each lsh code $\beta_\ell$, the reliability indice is 
directly based on the magnitude values of the $z_k$'s, for the bits which 
compose $\beta_\ell$. 

First, a reliability indice is computed for each component of $z_k$: 
considering the effect of degradation as an additive perturbation $e$ 
modeled by a Gaussian process, centered and with an empirical variance, 
we can compute the probability $p_k$ that $z_k+e$ changes sign. 
Then, according to the bit selection of each lsh function, 
we can derive the probability $\pi_\ell$ of a lsh code that 
at least one bit changes value, and $1-\pi_\ell$ is used as reliability indice 
for the lsh code $\beta_\ell$. 
Then, for each times and each band, the $L'<L$ lsh codes with the highest 
reliability are used to fill the hash table. 

Note only this makes possible to have a smaller hash table, with a shrinking 
factor $\sigma = L'/L$, but also the ratio $\rho = \mu / \nu$ increases, 
see eqs. (\ref{eq_mu},\ref{eq_nu}). 
The mean of code collisions $\nu$ with unmatching audio keys is reduced 
proportionally to $\sigma$, but the mean of good detections $\mu$ 
decreases significantly slower because the more reliable lsh codes are used, 
that is the corruption probability of the really used bits is lower. 
In the experiments, we chose $L'=10$, cf. sec.~\ref{sec-results}.

\subsection{Search step}\label{sec-search}

\begin{figure}[h]
  \begin{center}
    \includegraphics[width=\textwidth]{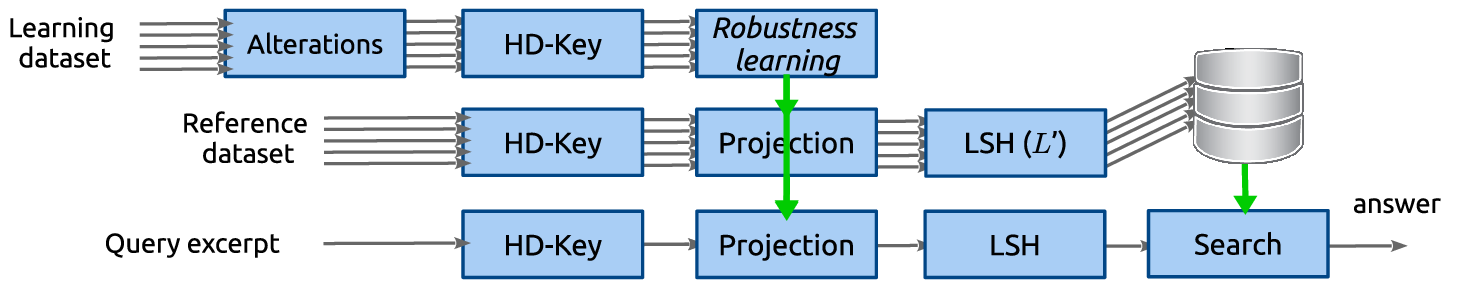}
  \end{center} 
\vspace*{-0.80cm}  
  \caption{\protect\label{fig-resume_all} 
    Illustration of the whole process. 
    First: the elements of the training dataset are altered, the corresponding
      HD-audio keys are computed, and the dimension reduction is learnt (see sec.~\ref{sec-DIAPA}). 
    Second: the HD-audio keys of the reference elements are computed, then 
      the dimension is reduced with the learnt reduction and the hash table is built.   
    Third: the HD-audio keys of the query excerpt are computed, and projected 
      with the same transformation, and all the $L$ lsh codes are used 
      to find matching codes in the hash table.}
\end{figure}

After having built the hash table with the music catalog of reference
items, see previous sub-sections, 
the lsh codes of the query excerpts are computed. 
Then, the number of matching codes between the query excerpt and 
each reference item of the catalog is fastly obtained using the hash tables. 
See an illustration of the process in Fig.~\ref{fig-resume_all}. 

Remark that during the search step, all the $L$ lsh codes obtained 
from the query excerpt are used. The selection of the $L'$ most reliable 
codes is applied only for the reference music piece during the 
building of the hash table, to reduce its size. 
The reliability indice of the lsh codes of the query excerpt 
can be used as weight, in order to favor the most robust codes. 

\subsubsection{Collision with unmatching items}

Nevertheless, collisions are possible with unmatching reference items. 
Indeed, the same hash code can be produced from different initial signals, 
and so, selecting the reference item only based on 
the number of matching codes is not enough, and may provide a high number 
of false positives. 
Note that because the audio prints are based on new audio descriptors which 
are relevant to the content of the input music, some times, the code 
collisions reveals music similarity, but it is not the case all the time. 

For example, with $b=16$ bits lsh codes, $L=51$ lsh functions, $L'=10$ 
reliable codes, $n_b=5$ bands, $F_a=4$ the mean number of analysis times 
per second, and 
for a $D_r=30$ sec long reference song, and 
    a $D_e=30$ sec long query excerpt, 
considering independent random drawing of lsh codes, 
the statistical mean of number of collisions 
between the query and the reference (with possible repetitions) is:  
\begin{equation} \label{eq_N_c}
  N_c = D_r D_e F_a^2 L' n_b/ 2^b \approx 10.98. 
\end{equation}
And in the case of the true matching reference without degradation, 
the number of good code matches (without repetition) is ideally: 
\begin{equation} \label{eq_N_t}
   N_t = \min(D_r, D_e) ~ F_a L' n_b = 6\,000. 
\end{equation}
But with degradations, this number strongly decreases, see Tab.~\ref{tab-tableK}.
For this reason, a post-process is applied, and it is based 
on the coherence of time positions of the matching codes between 
the query and some reference items. 
In sec.~\ref{sec-PIE} an overview of the time-coherence analysis is done. 

\subsubsection{Global search strategy}\label{ssec-hashSearch}

But analyzing the time-coherence with all the reference songs of the 
catalog would be too long. 
The used strategy approximately follows the approach of \cite{Wang2003}: 
\begin{itemize}
  \item[1] The hash table is used to select a small number $N$ of candidates, 
  \item[2] The time-coherence between the excerpt and the reference candidates 
            is analyzed in order to select the best candidate. 
\end{itemize}

Remark that, because the number of collisions is proportional 
to the duration $D_r$ of the reference song, see eq. (\ref{eq_N_c}), 
some problems may occur if some reference elements are significantly longer 
than others. 
A solution is to split the reference songs into short segments (e.g. 15sec), 
without overlap, and the segments are indexed rather than the complete songs. 
Then, the count of code matches per short segment is obtained through the 
hash table, 
and finally, this count is integrated 
on a longer sliding window, with overlap, and with a duration based on 
the query excerpt.

\subsubsection{Time-coherence analysis}\label{sec-PIE}

In this section, we briefly present the computation of a time coherence 
measure of matching codes between the query excerpt and a given 
candidate reference song. The selection of the candidate is 
presented in sec.~\ref{ssec-hashSearch}. 


In the initial work done in 2015, we adapted and improved the approach 
of \cite{Wang2003}. 
Given two sequences of time positions of matching codes, 
$t_n$ for the reference song and $\tau_n$ for the excerpt signal, 
the principle is to detect a linear relation for 
a subset of pairs $[t_n, \tau_n]$ in the plan $(t, \tau)$.  
Indeed, if the query is really an excerpt of the reference item, 
we may observe a linear alignment of some pairs $[t_n, \tau_n]$, 
with the relation $\tau_n = t_n - \delta_t^*$, with $\delta_t^*$ the starting 
time of the excerpt. 
But if it is not an excerpt of the reference song, the pairs 
correspond to false positives and should be randomly positioned 
in the plan ($t, \tau$). 

\begin{figure}[h]
  \begin{center}
    \includegraphics[width=\textwidth]{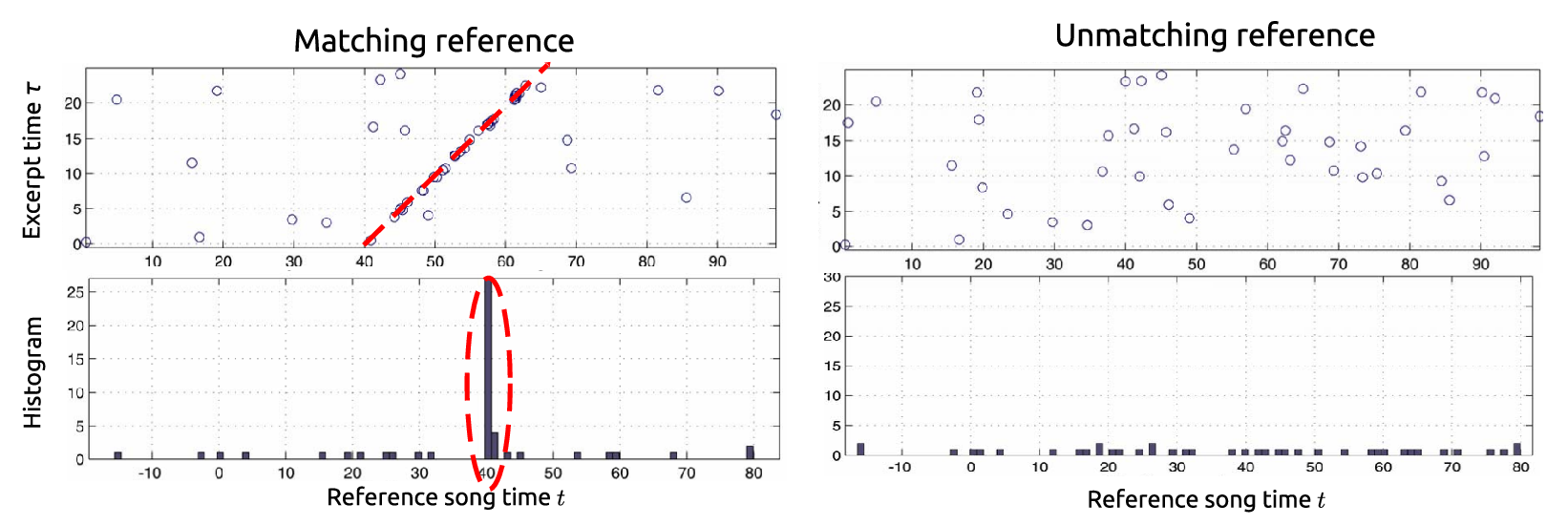}
  \end{center}    
\vspace*{-0.40cm}  
  \caption{\protect\label{fig-ShazamHistogram} 
    Illustration of the Shazam search strategy. 
    On the upper-left sub-figure are plotted all time pairs of matching 
    codes. The time coherence is displayed by the linear arrangement 
    around the red dashed line. The count permits to exhibit this matching, 
    see the bottom-left sub-figure. On the right sub-figures, 
    the reference item does not match with the input excerpt, the pairs 
    are random positioned in the plan $(t, \tau)$, and no linear arrangement 
    are observed. These figures come from \cite{Wang2003}.}
\end{figure}

To summarize the approach of \cite{Wang2003}, 
defining a short window size $\sigma$ (used as error margin), 
for different values of $\delta_t$ the number of pairs respecting 
$\| \delta_t + \tau_n -t_n \| \leq \sigma$ is count. 
Finally if a significant peak appear for a value of $\delta_t$, 
it is assume that the input excerpt matches with the tested reference item, 
and the obtained time offset is the value of $\delta_t^*$. 
Also, the count value at the found $\delta_t$ can be used 
as time-coherence measure to select or reject the candidates. 
An illustration of the search is presented in Fig.~\ref{fig-ShazamHistogram}. 
Remark that the computation is actually done in another plan given by the 
change of variables: ($t-\tau, \tau$). In this plan, the time coherence 
of pairs appears with a vertical alignment. This actually simplifies 
the calculus. 

Nevertheless, this approach does not deal with time stretching. 
In this case, we must consider the following relation: 
\begin{equation}
  \tau_n = \alpha t_n - \delta_t^*, 
\end{equation}
where $\alpha$ is the unknown time stretching factor. 
Applying the Shazam's method with time stretching may work with $\alpha$ 
close to 1, but the performances significantly fall with a stronger time 
stretching. 
A naive idea would to increase the margin $\sigma$ in order to cover 
the deviation due to the time stretching, but this is not sufficient. 

During the \emph{BeeMusic} project, in 2015, we proposed a pre-processing 
in order to significantly improve the result with 
a stronger time stretching. 
First, for each pair $n$ a cone is defined with 
($t_n, \tau_n$) as top, and delimited by two lines with slopes 
$\alpha_{\text{max}}$ and $1/\alpha_{\text{max}}$, 
where $\alpha_{\text{max}}>1$ is the maximal acceptable 
time stretching factor. 
Second, for each pair $n$, the number of other pairs inside the defined cone 
is used as weight. This weight is then used during the histogram count 
previously presented. 
In the case where the margin $\sigma$ is too small, 
because of the possible time stretching 
it may be possible that with the suited $\delta_t$, the integration windows 
are too short to count all the matching pairs of the line. 
Using this pre-processing, even if the window is too short, the
pairs on the line that are outside the window
are still taken into account through the weights of the pairs found inside.
Not only this pre-processing permits the analysis with time-stretching, 
but it also improves the results, without time-stretching, 
and with a true positive rate lower. 

Finally, 
a linear regression is computed with selected pairs 
in order to refine the estimation of the time stretching factor $\alpha$
and the time offset $\delta_t^*$. 


\renewcommand{\thesubsection}{\thesection.\arabic{subsection}}
\section{Evaluation and results}\label{sec-results}

This section briefly presents the evaluation of the performance of 
the recognition of a degraded music excerpt with a relatively big 
catalog of reference music. More details are presented in 
\cite{RemiIRCAMID_eval}, in french.

\subsection{Tested catalog}\label{ssec-dataset}

The tested catalog is based on a private dataset provided by a partner of the project, 
and internally usable for the period of the project. 
We extracted a subset with 755\,762 music excerpt with a duration 
between 25 and 35 seconds, and we removed duplicates based on the given 
ISRCs. 
A lot of different music genres are represented: 
ambiance, blues, classical, country, techno/dance, children, jazz, metal, 
world, soul/funk, rock, movie, pop, variety, rap/hip hop, raggae, gospel. 

For the first step of the search, the hash table uses 16 GigaBytes 
which can be loaded into the RAM. 
For the second step, all the data of the \emph{direct table} can be 
loaded into the RAM too, because it uses only 4 GigaBytes. 

\subsection{Two steps search}\label{ssec-2stepssearch}

For the recognition of a music excerpt, we present two measures of 
performances in Tab.~\ref{tab_results}. 
Each measure corresponds to a step in the search process. 

\paragraph{Step 1}~\\
The first displayed performance is the rate of good detections for step 1. 
For this step, a good detection happens when the 
reference item with the higher number of maching codes 
is the true reference song, which corresponds to the input excerpt.

\paragraph{Step 2}~\\
For the analysis of the time coherence, we finally adopted a different strategy
to define the number of candidates to test: 
with $N_i$ the number of code matches of the song $i$, sorted with decreasing 
$N_i$, so $N_1$ is the maximal value, the tested candidates $i$ are such that 
$N_i \geq N_1/2$. 
Nevertheless, we use a minimal number of 10 and a maximal number of 500 
candidates to test. 
Finally the best candidate is selected through the measure of the time-coherence, 
and the second displayed value in Tab.~\ref{tab_results} is the 
rate a good detection after the second step. 

\subsection{Evaluation protocol}\label{ssec-evalproto}

The hash table (step1) and the direct table (step2) are built with 
the 755\,762 reference excerpts (~30 seconds) without any degradations. 
Then we selected 7,146 test excerpts with 7 seconds duration from 
the reference catalog. 
Each tested excerpt is then degraded with 18 different degradations and 
with 3 different levels of degradations (strengths). 
Here are the tested degradations: 
\begin{itemize}
  \item\underline{\bf Graphical equalizers} with 10 bands, and with alternative gain: 
    $\pm \alpha$dB. The values of $\alpha$ for each of the 3 levels are: 
      {\bf 3dB, 6dB and 9dB}. Note that for the strongest degradation a lag of 18db 
      are between maxima and minima of the response. 
  \item \underline{{\bf Environmental noise} recorded in a \bf restaurant}. 
      The used SNRs (Signal-to-Noise Ratio) for the levels are: 
        {\bf 12dB, 6dB et 0dB}. Remark that for the strongest noise, its energy is the 
        same than the energy of the music to recognize. 
  \item \underline{{\bf Environmental noise }recorded in a \bf bus}. The same SNRs are used. 
  \item \underline{{\bf Environmental noise }recorded in a \bf street}. The same SNRs are used. 
  \item \underline{\bf Synthetic white noise}. The same SNRs are used. 
  \item \underline{\bf Synthetic pink noise}. The same SNRs are used.
  \item \underline{{\bf Pitch shifting to higher }frequencies}. 
        The software SuperVP is used here, with the preservation of the formants 
        and of the attack transients. The used transpositions are: 
        {\bf 1/4, 1/2, and 1 tone}. 
  \item \underline{{\bf Pitch shifting to lower }frequencies}. The used transpositions are: 
        {\bf -1/4, -1/2, and -1 tone}. 
  \item \underline{\bf Expansion of the time}: {\bf slow-down }time stretching. SuperVP is also used   
        with the same options. The used stretching are: 
        {\bf 15, 30 end 45 cents}. Remark, with $\beta$ the stretching in cents, 
        the duration is multiplied by the factor $2^{\beta/100}$. 
  \item \underline{\bf Compression of the time}: {\bf speed-up} time stretching. 
        The used stretching are: {\bf \mbox{-15,} -30 and -45 cents}.
  \item \underline{\bf MP3 coding}. The coding is done with the lame software and with the 
        following bitrates: {\bf 32kbps, 24kbps, and 16kbps}.
  \item \underline{\bf Distortion}. Here a $\arctan$ non-linear function is applied on the 
        time signal. With $L_\infty$ normalized input signal, the 
        level is given by the input gain (before the function): 
        {\bf 5dB, 12dB and 24dB}.
  \item \underline{\bf Dynamic range compressor}. We use here a 4 frequency bands compressor. 
        The level is both controlled by the compression ratio 
        and the release time: {\bf ratio 2, 8 and 50; release time 100ms, 10ms, et 1ms. }
  \item \underline{\bf Tremolo}. A cyclic modulation of amplitude (4Hz) is applied 
        with a gain lag (between maxima and minima): {\bf $\pm$3db, $\pm$6db, $\pm$9db}. 
  \item \underline{\bf Reverberation}. We use here a convolution reverberation 
        with an impulse response of a medium hall. The level is controled 
        with the mix ratio between the dry signal and the reverberated sound: 
        {\bf 9dB, 3dB, and 0dB}. 
  \item \underline{\bf Scenario GSM}.   
        Here we test a series of different degradations:  
        filtering with the response of a smartphone microphone (NexusOne), 
        white noise addition (SNRs {\bf 18dB, 12dB and 6dB}), 
        and a GSM coding is the main degradation (no parameter). 
\item   \underline{\bf Scenario slow-down}. 
        Here the following series of alterations is used: 
        slow-down with stretching factors: {\bf 4, 8 and 12 cents},  
        10-bands EQ with $\pm$3dB, 4-bands compressor with a ratio 2, 
        MP3 coding at 32kbps, reverberation with 3dB, 
        and noise addition with a 18dB SNR. 
\item   \underline{\bf Scenario noise}. 
        Here the following series of alterations is used: 
        slow-down (4\%), 10-bands EQ ($\pm$3dB), 4-bands compressor (ratio 2), 
        MP3 coding (32kbps), reverberation (3dB) and noise addition with SNRs:  
        {\bf 18dB, 12dB and 6dB}.
\end{itemize}

\subsection{Results}\label{ssec-finaleresults}

Table \ref{tab_results} gives the success rates for the 2 steps: 
\begin{itemize}
  \item[1)] \emph{STEP1:} ratio in percents of successful recognition 
               at the step 1, based on the hash table only   
                (with 755\,762 references). 
               This performance does not take into account the 
                time coherence analysis. 
  \item[2)] \emph{STEP2:} ratio in percents of successful recognition 
                at the step 2, based on the time-coherence analysis of the 
                best candidates selected after step 1. 
\end{itemize}

As expected the performances with the second step is better than 
with only the first first. 
Even if the performance of STEP1 is promising, 
this observation proves the benefit of the time coherence analysis. 
Indeed, in many cases the true reference song is among the selected 
candidates, but not the first. Others have a high number of matching codes,
but with random time positions, which do not fit a linear alignment. 
Unfortunately, in some cases, the true candidate is not among 
the selected candidates, and STEP2 cannot find it. 

This evaluation demonstrate the ability of the method to efficiently recognize 
a degraded excerpt among a big catalog, with more than 700\,000 music pieces, 
and with many different kind of degradations.  
But for some degradations, such as the noise addition, 
we know that this method is not as performant as other methods based 
on fingerprints, such as Shazam (spectrum peaks). 
Nevertheless, here other transformations such time stretching and pitch shifting
are managed without affecting the robustness to other degradations. 

For more details about this evaluation, other performances 
measures are presented in \cite{RemiIRCAMID_eval}. 

\begin{table}[h]
\centering
\makebox[0pt][c]{\parbox{1.\textwidth}{%
\hfill
    \begin{minipage}[b]{0.48\hsize}
      \centering
 \begin{tabular}{c||c|c|}
   &     \textbf{STEP 1} &   \textbf{STEP 2}   \\  
 \hline \multicolumn{3}{l}{ {\color{white}\huge ' }\textbf{Background noise: Restaurant}} \\ \hline  
strength 1 & \textbf{98.5} \% & \textbf{98.6} \%  \\  
 strength 2 & \textbf{96.8} \% & \textbf{97.0} \%  \\  
 strength 3 & \textbf{65.4} \% & \textbf{68.8} \%  \\  
 \hline \multicolumn{3}{l}{ {\color{white}\huge ' }\textbf{Background noise: Bus}} \\ \hline  
strength 1 & \textbf{98.5} \% & \textbf{98.6} \%  \\  
 strength 2 & \textbf{96.7} \% & \textbf{97.4} \%  \\  
 strength 3 & \textbf{70.9} \% & \textbf{81.0} \%  \\  
 \hline \multicolumn{3}{l}{ {\color{white}\huge ' }\textbf{Background noise: Street}} \\ \hline  
strength 1 & \textbf{98.5} \% & \textbf{98.5} \%  \\  
 strength 2 & \textbf{96.5} \% & \textbf{97.0} \%  \\  
 strength 3 & \textbf{67.5} \% & \textbf{71.5} \%  \\  
 \hline \multicolumn{3}{l}{ {\color{white}\huge ' }\textbf{White noise}} \\ \hline  
strength 1 & \textbf{98.7} \% & \textbf{98.9} \%  \\  
 strength 2 & \textbf{98.4} \% & \textbf{98.5} \%  \\  
 strength 3 & \textbf{92.3} \% & \textbf{95.8} \%  \\  
 \hline \multicolumn{3}{l}{ {\color{white}\huge ' }\textbf{Pink noise}} \\ \hline  
strength 1 & \textbf{98.6} \% & \textbf{98.7} \%  \\  
 strength 2 & \textbf{96.9} \% & \textbf{97.6} \%  \\  
 strength 3 & \textbf{74.5} \% & \textbf{84.5} \%  \\  
 \hline \multicolumn{3}{l}{ {\color{white}\huge ' }\textbf{Pitch Shifting plus}} \\ \hline  
strength 1 & \textbf{98.1} \% & \textbf{97.9} \%  \\  
 strength 2 & \textbf{95.5} \% & \textbf{97.0} \%  \\  
 strength 3 & \textbf{75.2} \% & \textbf{88.2} \%  \\  
 \hline \multicolumn{3}{l}{ {\color{white}\huge ' }\textbf{Pitch Shifting minus}} \\ \hline  
strength 1 & \textbf{98.2} \% & \textbf{98.1} \%  \\  
 strength 2 & \textbf{95.9} \% & \textbf{97.2} \%  \\  
 strength 3 & \textbf{77.4} \% & \textbf{89.2} \%  \\  
 \hline \multicolumn{3}{l}{ {\color{white}\huge ' }\textbf{Time Stretching slower}} \\ \hline  
strength 1 & \textbf{97.3} \% & \textbf{97.4} \%  \\  
 strength 2 & \textbf{93.5} \% & \textbf{95.4} \%  \\  
 strength 3 & \textbf{82.5} \% & \textbf{86.4} \%  \\  
 \hline \multicolumn{3}{l}{ {\color{white}\huge ' }\textbf{Time Stretching faster}} \\ \hline  
strength 1 & \textbf{97.6} \% & \textbf{97.8} \%  \\  
 strength 2 & \textbf{95.3} \% & \textbf{96.1} \%  \\  
 strength 3 & \textbf{91.5} \% & \textbf{87.8} \%  \\  
 \hline   
 \end{tabular} 
   \end{minipage}
    \hfill
 \begin{minipage}[b]{0.48\hsize}
  \centering
 \begin{tabular}{c||c|c|}
   &     \textbf{STEP 1} &   \textbf{STEP 2}   \\  
 \hline \multicolumn{3}{l}{ {\color{white}\huge ' }\textbf{Equalizer}} \\ \hline  
strength 1 & \textbf{98.7} \% & \textbf{98.7} \%  \\  
 strength 2 & \textbf{96.4} \% & \textbf{97.7} \%  \\  
 strength 3 & \textbf{71.2} \% & \textbf{87.9} \%  \\  
 \hline \multicolumn{3}{l}{ {\color{white}\huge ' }\textbf{MP3}} \\ \hline  
strength 1 & \textbf{98.9} \% & \textbf{98.8} \%  \\  
 strength 2 & \textbf{98.8} \% & \textbf{98.8} \%  \\  
 strength 3 & \textbf{98.5} \% & \textbf{98.5} \%  \\  
 \hline \multicolumn{3}{l}{ {\color{white}\huge ' }\textbf{Distortion}} \\ \hline  
strength 1 & \textbf{98.0} \% & \textbf{98.0} \%  \\  
 strength 2 & \textbf{95.9} \% & \textbf{96.9} \%  \\  
 strength 3 & \textbf{82.4} \% & \textbf{89.0} \%  \\  
 \hline \multicolumn{3}{l}{ {\color{white}\huge ' }\textbf{Multi-band Compressor}} \\ \hline  
strength 1 & \textbf{98.5} \% & \textbf{98.5} \%  \\  
 strength 2 & \textbf{92.2} \% & \textbf{95.5} \%  \\  
 strength 3 & \textbf{80.4} \% & \textbf{88.1} \%  \\  
 \hline \multicolumn{3}{l}{ {\color{white}\huge ' }\textbf{Tremolo}} \\ \hline  
strength 1 & \textbf{98.7} \% & \textbf{98.7} \%  \\  
 strength 2 & \textbf{97.9} \% & \textbf{98.0} \%  \\  
 strength 3 & \textbf{92.6} \% & \textbf{95.4} \%  \\  
 \hline \multicolumn{3}{l}{ {\color{white}\huge ' }\textbf{Reverberation}} \\ \hline  
strength 1 & \textbf{98.8} \% & \textbf{98.7} \%  \\  
 strength 2 & \textbf{97.5} \% & \textbf{97.8} \%  \\  
 strength 3 & \textbf{88.3} \% & \textbf{95.8} \%  \\  
 \hline \multicolumn{3}{l}{ {\color{white}\huge ' }\textbf{Scenario GSM}} \\ \hline  
strength 1 & \textbf{88.6} \% & \textbf{95.3} \%  \\  
 strength 2 & \textbf{86.3} \% & \textbf{94.0} \%  \\  
 strength 3 & \textbf{77.2} \% & \textbf{90.9} \%  \\  
 \hline \multicolumn{3}{l}{ {\color{white}\huge ' }\textbf{Scenario speed}} \\ \hline  
strength 1 & \textbf{93.5} \% & \textbf{96.6} \%  \\  
 strength 2 & \textbf{83.1} \% & \textbf{93.0} \%  \\  
 strength 3 & \textbf{61.9} \% & \textbf{80.3} \%  \\  
 \hline \multicolumn{3}{l}{ {\color{white}\huge ' }
  \textbf{Scenario noise~~~~~~~~~~~~~~~~~~~~~}} \\ \hline  
strength 1 & \textbf{93.6} \% & \textbf{96.5} \%  \\  
 strength 2 & \textbf{90.3} \% & \textbf{95.2} \%  \\  
 strength 3 & \textbf{76.6} \% & \textbf{86.6} \%  \\  
 \hline   
 \end{tabular} 
 \end{minipage}%
  \hfill
}}
\caption{Table of results: for all the 18 tested degradations, 
          with 3 different strenghts of degradations, and 
          the 2 steps. As expected, the performance 
          of the step 2 is higher than with step 1 only.}\label{tab_results}
\end{table}

\section{Conclusion}\label{sec-conclusion}

We presented in this report the achieved work which has the goal to 
improve the audio indexing system at IRCAM, with both a robustness 
to audio degradations and the feasibility do scale the method for big catalogs. 
The audio degradations are taken into account at different steps on the process: 
selection of the analysis times, sec.~\ref{sec-anchoredpoints}, 
high-dimensional audio prints with natural robustness to some 
degradation, sec.~\ref{sec-HDKey}, 
learning of dimension reduction of criteria based on robustness 
and discrimination for different initial signals, 
and with a preparation for the hashing, sec.~\ref{sec-DIAPA}, 
approximative hashing for a higher robustness to bit corruption, 
and so to audio degradations, sec.~\ref{sec-LSH}, 
and finally the search process based on the hash table and then 
a time-coherence analysis adapted for time stretching, sec.~\ref{sec-search}. 
The evaluation in sec.~\ref{sec-results} proves an interesting 
robustness and also the scalability of the method.

This work was achieved in 2015, in consequence, it did not take 
benefit of the recent advances in signal processing and machine learning. 
For example, it would be interesting to test 
the \emph{Scattering Transform} of \cite{anden2015joint,Lostanlen15}
in place of the HD-audio prints; with a possible adaptation to 
invariance to time and frequency compression/expansion rather 
than translation. 
Also, during the past years, significant progresses have been obtained 
in \emph{Deep Learning}, and keeping the same architecture of the present 
audio indexing system, we could improve many steps with \emph{Deep models}. 
For example, the analysis times selection should be easily implemented 
with a \emph{Neural Network}, where the ground truth is 
the output of the model on original signal. 
Moreover, the LDA and the OMPCA reductions used in this work 
are similar to a metric learning, and it should be possible to 
reimplement this reduction using a \emph{Triplet Loss} 
\cite{chechik2010large} for example,  
with additional losses related to the independences 
of the outputs variables, and the preparation for the 
hash table building.

\clearpage
\newpage

\subsection*{Acknowledgement}\label{Acknowledgement}
This work was funded by the \emph{BeeMusic} project (2013-2015), 
under the aegis of \emph{SNEP} and \emph{UPFI}, and with the 
coordination of \emph{Kantar Media}. 

\bibliography{refs}


~\cleardoublepage~
\thispagestyle{empty}

\end{document}